\documentclass[a4, 10pt, conference]{ieeeconf}      

\IEEEoverridecommandlockouts                              

\usepackage{graphics} 
\usepackage{epsfig} 
\usepackage{mathptmx} 
\usepackage{times} 
\usepackage{amsmath} 
\usepackage{amssymb}  
\usepackage{bm}
\usepackage{graphicx}
\usepackage{epstopdf}
\usepackage{supertabular}
\usepackage[usenames, dvipsnames]{color}
\usepackage{epstopdf}
\usepackage{arydshln}
\usepackage{caption}
\usepackage{verbatim}
\usepackage{cprotect}
\usepackage{mathtools}
\usepackage{subcaption}
\usepackage{url}
\usepackage[linesnumbered]{algorithm2e}

\usepackage{amsthm}
\usepackage{wrapfig}
\usepackage{float}
\usepackage[norsk]{babel}
\usepackage{multirow}

\theoremstyle{plain}

\def\i{\rm i}
\def\x{\rm x}

\def\e{\rm e}
\def\z{\rm z}
\def\y{\rm y}

\def\des{\rm des}
\def\rf{\rm ref}

\def\A{\rm A}
\def\t{\rm t}
\title{\LARGE \bf
Set-based Control for Autonomous Spray Painting
}

\author{Signe Moe$^{1}$, Jan T. Gravdahl$^{2}$ and Kristin Y. Pettersen$^{1}$ 
\thanks{$^{1}$S.Moe and K.Y.Pettersen are with the NTNU Center for Autonomous Marine Operations and Systems (NTNU AMOS), at The Department of Engineering Cybernetics, Norwegian University of Science and Technology (NTNU),
        Trondheim, Norway 
        {\tt\small $\left\lbrace \right.$signe.moe, kristin.y.pettersen$\left. \right\rbrace$@ntnu.no}}%
\thanks{$^{2}$J.T. Gravdahl is with The Department of Engineering Cybernetics, NTNU,
        Trondheim, Norway 
        {\tt\small jan.tommy.gravdahl@itk.ntnu.no}}
}

\begin{document}

\maketitle
\thispagestyle{empty}
\pagestyle{empty}

\begin{abstract}
In this paper, a method is presented for lowering the energy consumption and/or increasing the speed of a standard manipulator spray painting a surface. The approach is based on the observation that a small angle between the spray direction and the surface normal does not affect the quality of the paint job. Recent results in set-based kinematic control are utilized to develop a switched control system, where this angle is defined as a set-based task with a maximum allowed limit. Four different set-based methods are implemented and tested on a UR5 manipulator from Universal Robots. Experimental results verify the correctness of the method, and demonstrate that the set-based approaches can substantially lower the paint time and energy consumption compared to the current standard solution.
\end{abstract}

\section{INTRODUCTION}
Robotic systems are often required to perform one or several tasks which are given in the operational space, for instance obtaining a certain desired end effector position and/or orientation. However, these systems are often controlled in joint space, and thus a variety of inverse kinematics algorithms have been developed to map desired behavior from the operational space to the joint space and thus generate reference trajectories for the joint controllers. The most common approach is to use a Jacobian-based method~\cite{Caccavale2001}\nocite{Egeland1998}-\cite{Nenchev1992}, such as the Jacobian transpose, damped least squares or pseudo-inverse. In particular, the pseudo-inverse Jacobian is defined for systems that are not square nor have full rank and is a widely used solution to the inverse kinematics problem~\cite{Buss2004}\nocite{Klein1983}-\cite{Siciliano2008}.

A robotic system is said to be kinematically redundant if it possesses more degrees of freedom (DOFs) than those required to perform a certain task~\cite{Siciliano2009}. In this case, the ``excess'' DOFs can be utilized in order to perform several tasks simultaneously using the singularity-robust multiple task-priority inverse kinematics framework~\cite{Antonelli2008,Antonelli2009a}, which is widely used for a variety of applications~\cite{Antonelli1998}\nocite{Maciejewski1985,Shiyou2011,Tevatia2000}-\cite{Unicas}. This framework has been developed for \emph{equality tasks}, which specify exactly one desired value for certain given states of the system, for instance the position and orientation of the end effector. However, for a general robotic system, several goals may be described as \emph{set-based tasks}, which are tasks that have a desired interval of values rather than one exact desired value. Such tasks are also referred to as inequality constraints. Examples of such tasks are staying within joint limits~\cite{Marchand1996}, collision/obstacle avoidance~\cite{Hanafusa1981} and maintaining a high manipulability. 

A method to incorporate set-based tasks into the singularity-robust multiple task-priority inverse kinematics framework is presented in \cite{Antonelli2015,Moe2015} and is experimentally validated in \cite{Moe2015a}. A set-based task is ignored while the task value is within its valid set, and the remaining tasks of the system then decide the system trajectory. On the border of the valid set, the set-based task either remains ignored, or it is implemented as an equality task with the goal of freezing the task on the boundary. The proposed algorithm will choose the latter if the other tasks of the system push the set-based task out of its valid set. In the opposite case, the set-based task is still ignored. This results in a switched system with $2^n$ modes, where $n$ is the number of set-based tasks. 

Today, spray painting in manufacturing is mostly performed by robotic systems, and it is crucial that this task is performed both with high quality and in an efficient manner~\cite{From2010a}. In this paper, we represent the paint trajectory as a lawn mowing pattern defined by a radius of the turns, length of the straight line segments and initial position, but the proposed method is applicable also for other paths, which can be optimized for instance in terms of speed, coverage, and paint waste~\cite{Suh1991}\nocite{Kim2003a,Conner2005,Li2010a}-\cite{Tang2015}.

The standard method of spray painting consists of applying paint while keeping the spray nozzle normal to the spray surface. However, research shows that a small error in the orientation of the end effector relative to the surface does not affect the quality of the paint job. It is far more important to maintain constant velocity throughout the trajectory~\cite{From2007}. This is exploited in~\cite{From2010a}, where the nonlinear orientation constraints are transformed into positive definiteness constraints imposed on certain symmetric matrices to find the desired orientation every time-step. It is shown that this approach allows the end effector to maintain a higher constant velocity throughout the trajectory guaranteeing uniform paint coating and substantially reducing the time needed to paint the object, something that is experimentally validated in~\cite{From2011}.

In this paper, we suggest to define the angle between the paint nozzle and the spray surface as a high-priority set-based task in the control system. Thus, the entire spray process consists of one equality tasks (the spray task) in addition to the set-based orientation task, and the approach in~\cite{Moe2015} may be applied to generate reference velocities for the system joints. It has been proven that this method ensures satisfaction of the set-based tasks and asymptotic convergence of the equality task error given that the reference velocities are tracked and the tasks are compatible~\cite{Moe2016}. Unlike~\cite{From2010a}, where an optimization problem must be solved every time-step, the proposed method is deterministic and not dependent on fast convergence of numeric optimization solvers. Finally, in~\cite{From2010a}, the orientation of the spray nozzle is actively controlled throughout the entire operation, thereby occupying one or more DOFs at all times~\cite{From2010a}, whereas in the approach proposed in this paper, the orientation evolves freely according to the equality tasks until it is necessary to actively prevent the set-based task from being violated. Thus, the system has greater freedom to accomplish the spray task, which consists of \emph{the pointing task} (describes the point of intersection between the central axis of the end effector and the surface, i.e. the point where the paint would hit the surface~\cite{From2010a}) and \emph{the distance task} (the distance between the spray nozzle and the point of intersection). To ensure uniform coating, the latter is kept constant.

This paper is organized as follows. Section~\ref{sec:background} briefly described the singularity-robust task-priority inverse kinematics framework, upon which the proposed method is built, and an introduction to set-based control is given in Section~\ref{sec:set_based_control}. The proposed algorithm for a spray paint scenario is presented in Section~\ref{sec:set_based_spray}, and is validated by experimental results in Section~\ref{sec:experimental_results}. Finally, conclusions are given in Section~\ref{sec:conclusion}.

\section{THE SINGULARITY-ROBUST TASK-PRIORITY INVERSE KINEMATICS FRAMEWORK}
\label{sec:background}
This section presents the framework upon which set-based control is built.

A general robotic system has $n$ DOFs. Its configuration is given by the joint values $\bm{q} = [q_1, q_2,\dots,q_n]^T$. It is then possible to express tasks and task velocities in the operational space through forward kinematics and the task Jacobian matrix. The task variable that is to be controlled is given as $\bm{\sigma}(t)\!\in\!\mathbb{R}^m$,
\begin{equation}\label{eq:f}
   \bm{\sigma}(t) = \bm{f}(\bm{q}(t)),
\end{equation}
where $\bm{f}(\bm{q}(t))$ is the forward kinematics, which can be derived for instance through the Denavit-Hartenberg convention~\cite{Spong2005}. The time-derivative of the task is given as
\begin{equation}
   \dot{\bm{\sigma}}(t) = \frac{\partial\bm{f}(\bm{q}(t))}{\partial\bm{q}}\dot{\bm{q}}(t)= \bm{J}(\bm{q}(t))\dot{\bm{q}}(t),
   \label{eq:J}
\end{equation}
where $\bm{J}(\bm{q}(t))\!\in\!\mathbb{R}^{m\times n}$ is the configuration-dependent analytical task Jacobian matrix and ${\bm{\dot{q}}}(t)\!\in\!\mathbb{R}^n$ is the system velocity. For compactness, the argument $\bm{q}$ of tasks and Jacobians are omitted from the equations for the remainder of this section.

Consider a single $m$-dimensional task to be followed, with a defined desired trajectory ~$\bm{\sigma}_{\des}(t)\!\in\!\mathbb{R}^m$. The corresponding joint references $\bm{q}_{\des}(t)\!\in\!\mathbb{R}^n$ for the robotic system may be computed by integrating the locally inverse mapping of~\eqref{eq:J} achieved by imposing minimum-norm velocity. The following least-squares solution is given:
\begin{equation}\label{eq:Jinv}
    \dot{\bm{q}}_{\des} = \bm{J}^\dag \dot{\bm{\sigma}}_{\des} = \bm{J}^T\left(\bm{J}\bm{J}^T\right)^{-1} \dot{\bm{\sigma}}_{\des},
\end{equation}
where~$\bm{J}^\dag$, implicitly defined in the above equation for full  row  rank matrices, is the right pseudoinverse of~$\bm{J}$. In the general case, the pseudoinverse is the matrix that satisfies the four Moore-Penrose conditions (\ref{eq:MP_1})-(\ref{eq:MP_4})~\cite{Golub1996}, and it is defined for systems that are not square ($m \neq n$) nor have full rank~\cite{Buss2004}:

\begin{align}
\bm{JJ}^{\dag}\bm{J} &= \bm{J}, \label{eq:MP_1}\\
\bm{J}^{\dag}\bm{JJ}^{\dag} &= \bm{J}^{\dag}, \\
(\bm{JJ}^{\dag})^{\star} &= \bm{JJ}^{\dag}, \label{eq:hermitian1}\\
(\bm{J}^{\dag}\bm{J})^{\star} &= \bm{J}^{\dag}\bm{J}. \label{eq:MP_4}
\end{align}
Here, $\bm{J}^{\star}$ denotes the complex-conjugate of $\bm{J}$. 

The vector $\bm{q}_{\des}$, which is found by taking the time integral of~\eqref{eq:Jinv}, is prone to drifting. To handle this, a closed loop inverse kinematics (CLIK) version of the algorithm is usually implemented~\cite{Chiaverini1997}.
\begin{equation}\label{eq:CLIK}
    \dot{\bm{q}}_{\des} = \bm{J}^\dag\Bigl(\dot{\bm{\sigma}}_{\des}+ \bm{\Lambda}\bm{\tilde{\sigma}}\Bigr) = \bm{J}^\dag\dot{\bm{\sigma}}_{\rf},
\end{equation}
where $\tilde{\bm{\sigma}}\in\mathbb{R}^m$ is the task error defined as
\begin{equation}
    \tilde{\bm{\sigma}}=\bm{\sigma}_{\des}-\bm{\sigma}
\end{equation}
and $\bm{\Lambda}\in\mathbb{R}^{m\times m}$ is a positive-definite matrix of gains. This feedback approach reduces the error dynamics to
\begin{equation}
\begin{split}
\bm{\dot{\tilde{\sigma}}} &= \bm{\dot{\sigma}}_{\des}-\bm{\dot{\sigma}} = \bm{\dot{\sigma}}_{\des}-\bm{J}\bm{\dot{q}} \\
&= \bm{\dot{\sigma}}_{\des}-\bm{J}\bm{J}^\dag(\bm{\dot{\sigma}}_{\des}+\bm{\Lambda}\bm{\tilde{\sigma}}) \\ 
& = -\bm{\Lambda \tilde{\sigma}},
\end{split}
\label{eq:one_task}
\end{equation}
if $\bm{\dot{q}} = \bm{\dot{q}}_{\des}$ and $\bm{J}$ has full rank, implying that $\bm{J}\bm{J}^\dag = \bm{I}$. Equation~(\ref{eq:one_task}) describes a linear system with a  globally  exponentially stable equilibrium point at the equilibrium $\bm{\tilde{\sigma}} = \bm{0}$. It is worth noticing that the assumption $\bm{\dot{q}} = \bm{\dot{q}}_{\des}$ is common to all inverse kinematics algorithms. For practical applications, it requires that the low level dynamic control loop is faster than the kinematic one.

\section{SET-BASED CONTROL}
\label{sec:set_based_control}
This section presents the idea behind set-based control within the singularity-robust multiple task-priority inverse kinematics framework. This framework has been developed for equality tasks that have a specific desired value $\bm{\sigma}_{\des}(t)$, e.g.\ the desired end effector position. Set-based control is a method to extend the existing framework to handle set-based tasks such as the avoidance of joint limits and obstacles, field of view etc.

A set-based task is expressed through forward kinematics (\ref{eq:f}), but the objective is to keep the task in a defined set $D$ rather than controlling it to a desired value. Mathematically, this can be expressed as $\bm{\sigma}(t) \in D~\forall~t$ rather than $\bm{\sigma}(t) = \bm{\sigma}_{\des}(t)$. Thus, set-based tasks cannot be directly inserted into the singularity-robust multiple task-priority inverse kinematics framework.

\begin{figure}[htbp]
\centering
\includegraphics[width=0.8\columnwidth]{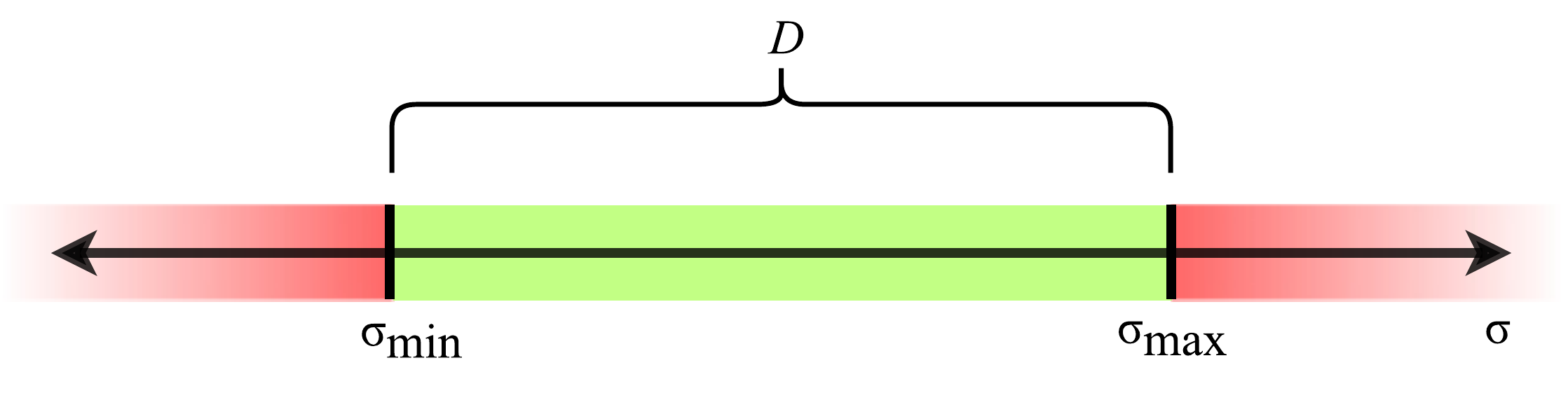}
\caption{Illustration of valid set $D$. The set-based task $\sigma$ is satisfied in $D$ and violated outside of $D$.}
\label{fig:valid_set_D}
\end{figure}

Consider Fig.~\ref{fig:valid_set_D}. A set-based task $\sigma$ is defined as \textit{satisfied} when it is contained in its valid set, i.e.\ $\sigma \in D = [\sigma_{\min},\sigma_{\max}]$. On the boundary of $D$ the task is still satisfied, but it might be necessary to actively handle the task to prevent it from becoming violated. 

\begin{figure}[H]
	\centering
	\includegraphics[width=0.45\textwidth]{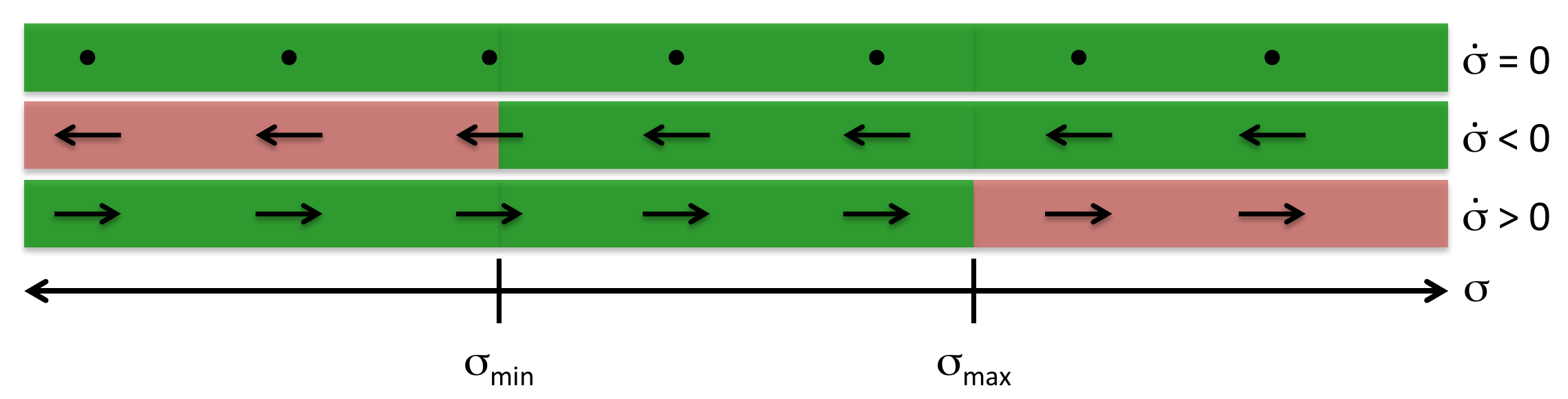}
	\caption{Graphic illustration of the tangent cone function in\_T\_C with return value \color{ForestGreen}True\color{black}~shown in green and \color{BrickRed}False\color{black}~in red}
	\label{fig:tangent_cone}
\end{figure}

In~\cite{Antonelli2015}\nocite{Moe2015}-\cite{Moe2016} a method is presented that allows a general number of scalar set-based tasks to be handled in this framework with a given priority within a number of equality tasks, and experimental results and practical implementation are given in~\cite{Moe2015a,Moe2016}. The results in this paper are based on these methods, and a crucial element to set-based control is the tangent cone to the set $D$, which is given as
\begin{equation}
T_{D}(\sigma) = \left\lbrace \begin{matrix} \left[\right.0,\infty \left.\right) & {\sigma} = {\sigma}_{\min} \\\mathbb{R} & {\sigma} \in ({\sigma}_{\min}, {\sigma}_{\max}) \\ \left( \right.-\infty,0\left.\right] & {\sigma} = {\sigma}_{\max} \end{matrix} \right..
\label{eq:tangent_cone_def}
\end{equation}
Note that if $\dot{\sigma}(t) \in T_D(\sigma(t))~\forall~t \ge t_0$, then this implies that $\sigma(t) \in~D~\forall~t \ge t_0$. If $\sigma$ is in the interior of $D$, the derivative is always in the tangent cone, as this is defined as $\mathbb{R}$. If $\sigma = \sigma_{\min}$, the task is at the lower border of the set. In this case, if $\dot{\sigma} \in [0,\infty)$, then $\sigma$ will either stay on the border, or move into the interior of the set. Similarly, if $\sigma = \sigma_{\max}$ and $\dot{\sigma} \in (-\infty,0]$, $\sigma$ will not leave $D$. Hence, the goal of set-based control is to define one or more tasks with corresponding valid sets, and ensure that the task derivatives are always contained in the tangent cone of said sets. The tangent cone implementation is given in Algorithm~\ref{alg:in_t_c} and illustrated in Figure~\ref{fig:tangent_cone}.

\begin{algorithm}
\KwIn{$\dot{\sigma}$, ${\sigma}$, $\sigma_{\min}$, $\sigma_{\max}$}
\uIf{$\sigma_{\min} < \sigma < \sigma_{\max}$}{\Return True\;}
\uElseIf{$\sigma \le \sigma_{\min}$ and $\dot{\sigma} \ge 0 $}{\Return True\;}
\uElseIf{$\sigma \le \sigma_{\min}$ and $\dot{\sigma} < 0 $}{\Return False\;}
\uElseIf{$\sigma \ge \sigma_{\max}$ and $\dot{\sigma} \le 0 $}{\Return True\;}
\Else{\Return False\;}
\caption{The boolean function in\_T\_C.}
\label{alg:in_t_c}
\end{algorithm}

\section{SET-BASED CONTROL OF SPRAY PAINT}
\label{sec:set_based_spray}
This section presents the implementation of set-based control for a spray paint scenario. Traditionally, paint is applied by a robotic system with a nozzle that is controlled to be perpendicular to the spray surface. However, it can be shown that the velocity of the paint gun is far more important than the orientation when it comes to uniform paint coating. A small orientation error ($\le 20^{\circ}$) in the paint gun does not affect the quality of the coating to the same extent as changes in the velocity~\cite{From2010a}. Based on this, the angle between the paint gun and the surface normal may be described as a set-based task.

\subsection{EXPERIMENTAL SETUP}
\begin{figure}[htb] 
\centering
\includegraphics[width=0.25\textwidth]{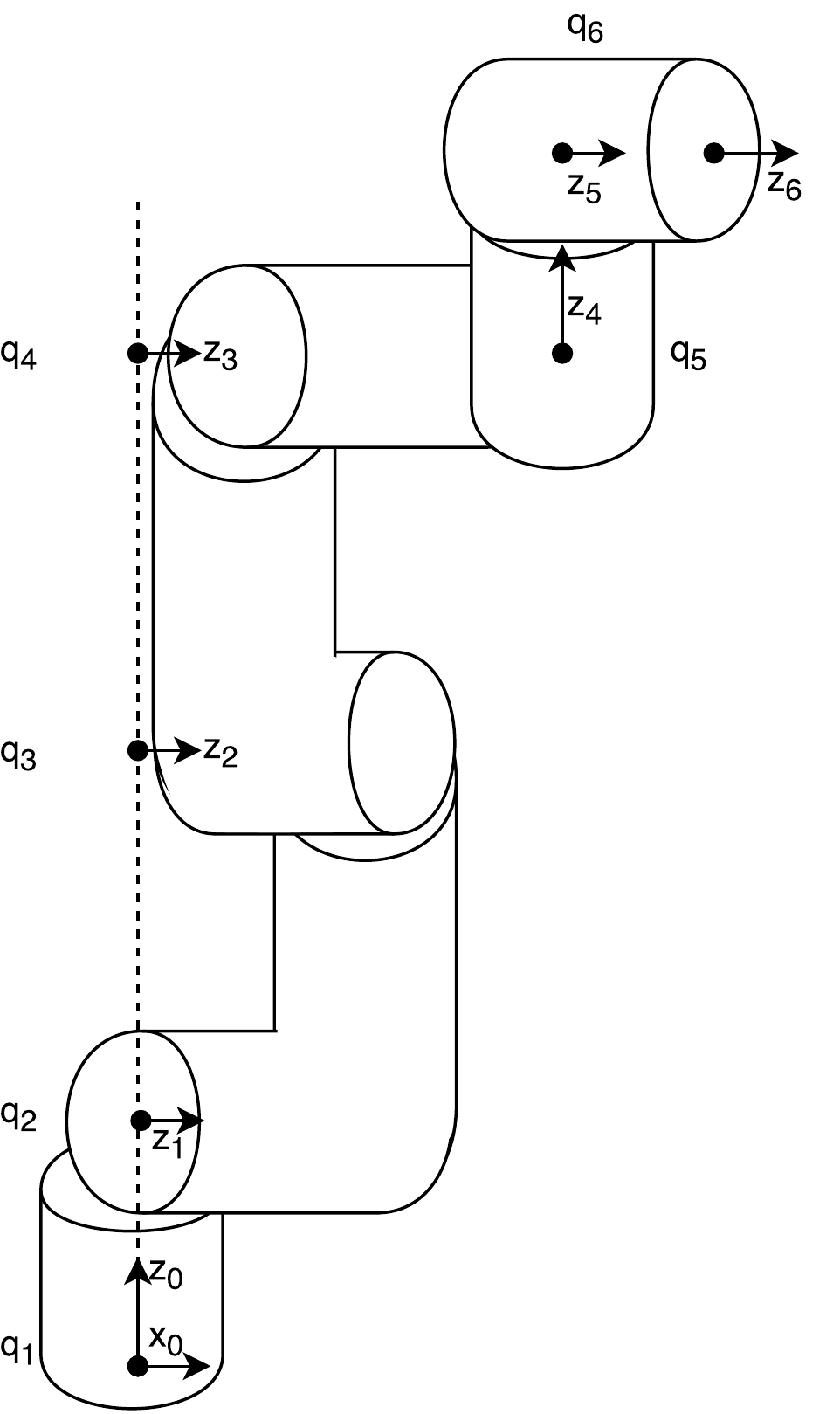}
\caption{Coordinate frames corresponding to the D-H parameters in Table~\ref{tab:DH}.}
\label{fig:ur5_coordinate}
\end{figure}

To explore the effects of set-based control for a spray paint scenario, experiments have been run on UR5 manipulator from Universal robots. The UR5 has 6 revolute joints, and the joint angles are denoted $\bm{q} \triangleq \begin{bmatrix}q_1 & q_2 & q_3 & q_4 & q_5 & q_6\end{bmatrix}^\textrm{T}$. The Denavit-Hartenberg (D-H) parameters are used to calculate the forward kinematics (i.e. the position and the orientation of the end effector as a function of $\bm{q}$). The parameters are given in~\cite{Wu2014} and are presented Table~\ref{tab:DH} with the corresponding coordinate frames illustrated in Fig.~\ref{fig:ur5_coordinate}. The resulting forward kinematics has been experimentally verified to confirm the correctness of the parameters. 

\begin{table}[H]
\centering
\begin{tabular}{|c|c|c|c|c|}
\hline
Joint & $a_i$ [m] & $\alpha_i$ [rad] & $d_i$ [m] & $\theta_i$ [rad] \\
\hline
1 & 0      & $\pi$/2   & 0.089 & $q_1$ \\
2 & -0.425 & 0         & 0     & $q_2$ \\
3 & -0.392 & 0         & 0     & $q_3$ \\
4 & 0      & $\pi$/2   & 0.109 & $q_4$ \\
5 & 0      & -$\pi$/2  & 0.095 & $q_5$ \\
6 & 0      & 0         & 0.082 & $q_6$ \\
\hline 
\end{tabular}
\caption{Table of the D-H parameters of the UR5. The corresponding coordinate systems can be seen in Fig.~\ref{fig:ur5_coordinate}.}
\label{tab:DH}
\end{table}

The UR5 is equipped with a high-level controller that can control the robot both in joint and Cartesian space. In the experiments presented here, a calculated reference $\bm{q}_{\textrm{des}}$ is sent to the high-level controller, which is assumed to function nominally such that
\begin{equation}
\begin{split}
\bm{q} &\approx \bm{q}_{\textrm{des}}.
\end{split}
\end{equation}
From this reference, $\bm{\dot{q}}_{\textrm{des}}$ and $\bm{\ddot{q}}_{\textrm{des}}$ are extrapolated and sent with $\bm{q}_\textrm{des}$ to the low-level controller.

\begin{figure}[htbp]
\centering
\includegraphics[width=0.48\textwidth]{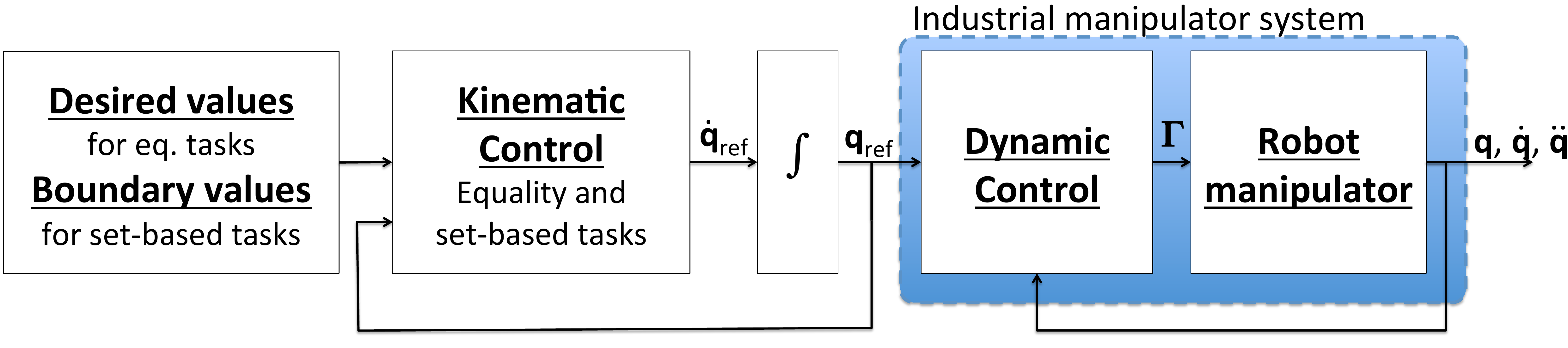}
\caption{The control structure of the experiments. The tested algorithm is implemented in the kinematic controller block.}
\label{fig:control_structure}
\end{figure}

The structure of the system is illustrated in Fig.~\ref{fig:control_structure}. The algorithm described in Section~\ref{sec:implementation} is implemented in the kinematic controller block. Every timestep, a reference for the joint velocities is calculated and integrated to desired joint angles $\bm{q}_{\textrm{des}}$. This is used as input to the dynamic controller, which in turn applies torques to the joint motors.  Note that the actual state $\bm{q}$ is not used for feedback to the kinematic control block. When the current state is used as input for the kinematic controller, the kinematic and dynamic loops are coupled and the gains designed for the kinematic control alone according to~\cite{Moe2016} can not be used. This results in uneven motion, and therefore the kinematic control block receives the previous reference as feedback, which leads to much nicer behavior, and is a good approximation because the dynamic controller tracks the reference with very high precision. This is the standard method of implementation for industry robots when kinematic control is used. 

The communication between the implemented algorithm and the industrial manipulator system occurs through a TCP/IP connection which operates at {$\text{125 Hz}$}.  The kinematic control block is implemented using python, which is a very suitable programming language for the task. The TCP/IP connection is very simple to set up in python. Furthermore, python has several libraries that can handle different math and matrix operations. 

\subsection{IMPLEMENTED TASKS}
\label{sec:implementes_task}
Several different tasks make up the basis for a spray paint scenario.

\subsubsection{Position control}
\begin{figure}
\centering
\includegraphics[width=0.7\linewidth]{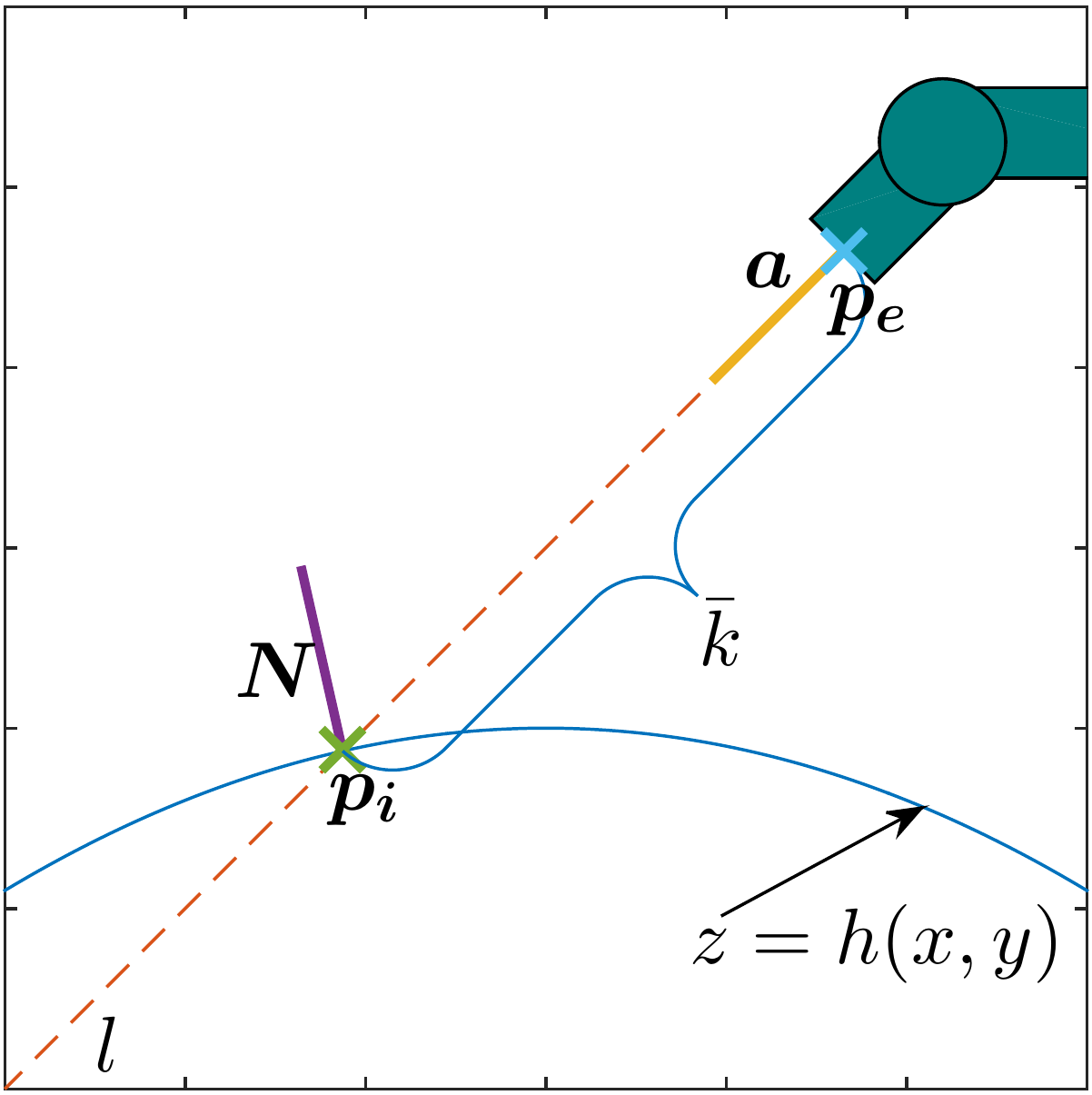}
\caption{Illustration of spray paint scenario.}
\label{fig:spray_variables}
\end{figure}

The position of the end effector $\bm{p}_{\e}= \begin{bmatrix} x_{\e} & y_{\e} & z_{\e}\end{bmatrix}^T$ (see Fig.~\ref{fig:spray_variables}) relative to the base coordinate frame is given by the forward kinematics. The analytical expression can be found through the homogeneous transformation matrix~\cite{Spong2005} using the D-H parameters given in Table~\ref{tab:DH}. The task is then defined by

\begin{align}
\bm{\sigma}_{\textrm{pos}}(\bm{q}) &= \bm{p}_{\e}(\bm{q}) = \bm{f}(\bm{q}) \in \mathbb{R}^3 \\
\bm{\dot{\sigma}}_{\textrm{pos}}(\bm{q},\bm{\dot{q}}) &= \bm{J}_{\textrm{pos}}(\bm{q})\bm{\dot{q}} = \frac{\text{d}\bm{f}}{\text{d}\bm{q}}\bm{\dot{q}},
\end{align}
\noindent where the function $\bm{f}(\bm{q})$ is given by the forward kinematics.

\subsubsection{Field of view}
The field of view is defined as the outgoing unit vector of the end effector, i.e. the $z_6$-axis in Figure~\ref{fig:ur5_coordinate}. This vector expressed in base coordinates is denoted $\bm{a}(\bm{q}) = \begin{bmatrix}a_{\x}(\bm{q}) & a_{\y}(\bm{q}) & a_{\z}(\bm{q})\end{bmatrix}^T \in \mathbb{R}^3$, and can be found through the homogeneous transformation matrix using the D-H parameters. 

It is very useful to control the FOV when directional devices, sensors or, as in this particular case, a spray nozzle are mounted on the end-effector and these are desired to point in a certain direction $\bm{a}_{\textrm{des}}(\bm{q}) \in \mathbb{R}^3$. We then define a one-dimensional FOV task as the norm of the error between $\bm{a}$ and $\bm{a}_{\textrm{des}}$:
\begin{align}
{\sigma}_{\textrm{FOV}}(\bm{q}) &= \sqrt{(\bm{a}_{\textrm{des}}(\bm{q})-\bm{a}(\bm{q}))^\textrm{T}(\bm{a}_{\textrm{des}}(\bm{q})-\bm{a}(\bm{q}))} \in \mathbb{R} \label{eq:sigma_FOV}\\
{\dot{\sigma}}_{\textrm{FOV}}(\bm{q},\bm{\dot{q}}) &= \bm{J}_{\textrm{FOV}}(\bm{q})\bm{\dot{q}} \nonumber \\
&= \frac{(\bm{a}_{\textrm{des}}(\bm{q})-\bm{a}(\bm{q}))^\textrm{T}}{\sigma_{\textrm{FOV}}(\bm{q})}\left(\frac{\text{d} \bm{a}_{\des}(\bm{q})}{\text{d} \bm{q}}-\frac{\text{d} \bm{a}(\bm{q})}{\text{d} \bm{q}}\right)\bm{\dot{q}} \label{eq:J_FOV}
\end{align}
Note that $\bm{J}_{\textrm{FOV}}$ is not defined for $\sigma_{\textrm{FOV}} = 0$. In the implementation, this is solved by adding a small $\epsilon>0$ to the denominator of this Jacobian, thereby ensuring that division by zero does not occur.

In these experiments, the vector $\bm{a}$ corresponds to the direction of the spray nozzle and is therefore highly relevant for a spray paint scenario. For a traditional spray paint scenario, $\bm{a}_{\des} = -\bm{N}$, where $\bm{N}$ is the normal to the spray surface. This is illustrated in Fig.~\ref{fig:spray_variables}.

\subsubsection{Spray task}
For spray painting, it is desirable to control the point of intersection $\bm{p}_{\i} \in \mathbb{R}^2$, which is the point where the spray hits the spray surface. This surface is described by the function $h(x,y)$ such that
\begin{equation}
z = h(x,y),
\end{equation}
and is illustrated in Fig.~\ref{fig:spray_variables}. The intersection point is given as the point the line $l$ intersects the surface $z = h(x,y)$. The line $l$ can be parametrized by $k\ge 0$ as
\begin{equation}
l(k,\bm{q}):= \bm{p}_{\e}(\bm{q}) + \bm{a}(\bm{q})k = \begin{bmatrix}
x_{\e}(\bm{q}) + a_x(\bm{q})k \\ y_{\e}(\bm{q}) + a_y(\bm{q})k \\ z_{\e}(\bm{q}) + a_z(\bm{q})k 
\end{bmatrix}_.\label{eq:line_l}
\end{equation}
The point of intersection occurs for $k=\bar{k}(\bm{q})$, which can be calculated by solving the equation
\begin{equation}
z_{\e}(\bm{q}) + a_z(\bm{q})\bar{k} = h(x_{\e}(\bm{q}) + a_x(\bm{q})\bar{k},y_{\e}(\bm{q}) + a_y(\bm{q})\bar{k}). \label{eq:bar_k}
\end{equation}
Note that depending on the surface function $h$, the above equation may not have an analytical solution for $\bar{k}$. For practical purposes $\bar{k}$ may be estimated by using optimization techniques.

\begin{figure}
\centering
\includegraphics[width=0.7\linewidth]{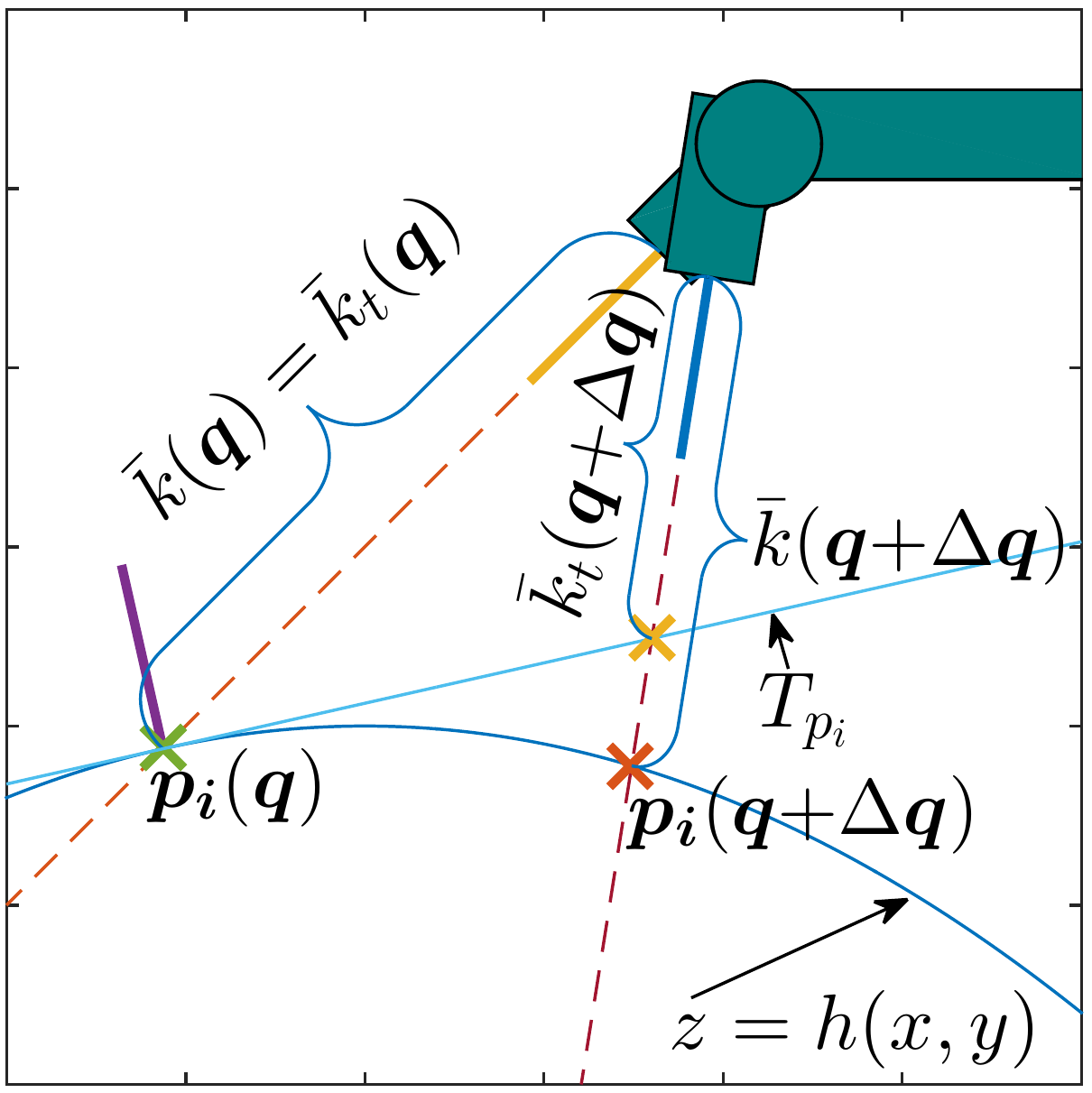}
\caption{Illustration of $\bar{k}$ and $\bar{k}_{\t}$.}
\label{fig:tangent_plane}
\end{figure}

In addition to controlling the point of intersection, it is necessary to control the distance between the surface and the spray gun, which is given by $\bar{k}$, to achieve a uniform paint coat. Since it may be challenging to find an explicit expression for $\bar{k}$, the task Jacobian is calculated by approximating the spray surface as the tangent plane in the intersection point. Hence, for implementation purposes, the following steps are taken: First, the numeric value for $\bar{k}$ and the resulting $\bm{p}_{\i}$ are calculated based on~(\ref{eq:bar_k}) using optimization techniques, and
\begin{align}
\bm{p}_{\i}(\bm{q}) &= \begin{bmatrix}
x_{\i}(\bm{q}) \\ y_{\i} (\bm{q})
\end{bmatrix} = \begin{bmatrix}
x_{\e}(\bm{q})\\y_{\e}(\bm{q})
\end{bmatrix} + \bar{k}(\bm{q})\begin{bmatrix}
a_{\x}(\bm{q}) \\ a_{\y}(\bm{q})
\end{bmatrix}_, \\
z_{\i}(\bm{q}) &= h\left(x_{\i}(\bm{q}),y_{\i}(\bm{q})\right).
\end{align}
Then, the normal vector $\bm{N}$ to the surface $z=h(x,y)$ in the point $(\bm{p}_{\i}, z_{\i})$ is found as
\begin{equation}
\bm{N}(\bm{q}) = \begin{bmatrix}
-\frac{\delta h}{\delta x}(x_{\i}(\bm{q}),y_{\i}(\bm{q})) \\ -\frac{\delta h}{\delta y}(x_{\i}(\bm{q}),y_{\i}(\bm{q})) \\ 1
\end{bmatrix} = \begin{bmatrix}
n_1 \\ n_2 \\ n_3
\end{bmatrix}_,
\end{equation}
and the tangent plane in the same point is given as
\begin{equation}
T_{\bm{p}_{\i}}: n_1(x-x_{\i}) + n_2(y-y_{\i}) + n_3(z-z_{\i}) = 0. \label{eq:tangent_plane}
\end{equation}
Note that even though the point $\bm{p}_{\i}$ and $\bm{N}$ change as a function of the robot configuration $\bm{q}$, the above equation describes the tangent plane in one specific $\bm{p}_{\i}$, and thus these values are constants. Although it may be challenging, or even impossible, to derive an explicit expression for the distance $\bar{k}$ between the end effector position $\bm{p}_{\e}$ and a general surface $z=h(x,y)$ along the line $l$, it is straight forward to calculate the same distance between the end effector position and the tangent plane. This distance is denoted $\bar{k}_{\t}$. To do so, we insert~(\ref{eq:line_l}) into~(\ref{eq:tangent_plane}), which yields
\begin{align}
\bar{k}_{\t}(\bm{q}) &= -\frac{n_1(x_{\e}(\bm{q})-x_{\i})+n_2(y_{\e}(\bm{q})-y_{\i})+n_3(z_{\e}(\bm{q})-z_{\i})}{n_1a_{\x}(\bm{q})+n_2a_{\y}(\bm{q})+n_3a_{\z}(\bm{q})}\nonumber \\
&= -\frac{a(\bm{q})}{b(\bm{q})}. \label{eq:bar_k_t}
\end{align}
Then, taking the time derivative of~(\ref{eq:bar_k_t}), it is clear that
\normalsize
\begin{align}
\dot{\bar{k}}_{\t}(\bm{q},\bm{\dot{q}}) &= -\frac{(n_1\dot{x}_{\e}(\bm{q},\bm{\dot{q}})+n_2\dot{y}_{\e}(\bm{q},\bm{\dot{q}})+n_3\dot{z}_{\e}(\bm{q},\bm{\dot{q}}))b(\bm{q})}{b(\bm{q})^2} \nonumber \\
&~~~+\frac{a(\bm{q})(n_1\dot{a}_{\x}(\bm{q},\bm{\dot{q}})+n_2\dot{a}_{\y}(\bm{q},\bm{\dot{q}})+n_3\dot{a}_{\z}(\bm{q},\bm{\dot{q}}))}{b(\bm{q})^2} \nonumber \\
&=-\frac{b(\bm{q})(n_1\bm{J}_{\textrm{pos}_1}(\bm{q})+n_2\bm{J}_{\textrm{pos}_2}(\bm{q})+n_3\bm{J}_{\textrm{pos}_3}(\bm{q}))}{b(\bm{q})^2}\bm{\dot{q}} \nonumber \\
&~~~+\frac{a(\bm{q})(n_1\bm{J}_{\textrm{FOV}_1}(\bm{q})+n_2\bm{J}_{\textrm{FOV}_2}(\bm{q})+n_3\bm{J}_{\textrm{FOV}_3}(\bm{q}))}{b(\bm{q})^2}\bm{\dot{q}} \nonumber \\
&= \bm{J}_{\textrm{dist}_t}(\bm{q})\bm{\dot{q}},
\end{align}
\normalsize

\noindent where $\bm{J}_{\A_i}$ denotes the $i$th row of the matrix $\bm{J}_{\A}$. For small changes $\bm{\dot{q}}$ (i.e. small time steps), the approximation
\begin{equation}
\dot{\bar{k}} \approx \dot{\bar{k}}_{\t} = \bm{J}_{\textrm{dist}_t}(\bm{q})\bm{\dot{q}}
\end{equation}
is sufficient. Thus, in the implementation, we estimate the corresponding change in $\bar{k}$ related to a change in configuration $\bm{\dot{q}}$ as the change in the distance between the end effector and the tangent plane $T_{\bm{p}_{\i}}$~(\ref{eq:tangent_plane}), see Fig.~\ref{fig:tangent_plane}. Note that this approach does not result in drifting over time, as each time step the exact $\bar{k}$ is found numerically and a new tangent plane is calculated.

Hence, we define the spray task $\bm{\sigma}_{\textrm{spray}} \in \mathbb{R}^3$ as
\begin{align}
\bm{\sigma}_{\textrm{spray}}(\bm{q}) &= \begin{bmatrix}
\bm{p}_{\i}(\bm{q}) \\ \bar{k}(\bm{q})
\end{bmatrix} = \begin{bmatrix}
x_{\e}(\bm{q}) + a_{\x}(\bm{q})\bar{k}(\bm{q}) \\ 
y_{\e}(\bm{q}) + a_{\y}(\bm{q})\bar{k}(\bm{q}) \\
\bar{k}(\bm{q})
\end{bmatrix}_,\label{eq:sigma_spray}\\
\end{align}
\begin{align}
\bm{\dot{\sigma}}_{\textrm{spray}}(\bm{q},\bm{\dot{q}}) &= \begin{bmatrix}
\dot{x}_{\e}(\bm{q},\bm{\dot{q}})+ \dot{a}_{\x}(\bm{q},\bm{\dot{q}})\bar{k}(\bm{q}) + a_{\x}(\bm{q})\dot{\bar{k}}(\bm{q},\bm{\dot{q}}) \\ \dot{y}_{\e}(\bm{q},\bm{\dot{q}})+ \dot{a}_{\y}(\bm{q},\bm{\dot{q}})\bar{k}(\bm{q}) + a_{\y}(\bm{q})\dot{\bar{k}}(\bm{q},\bm{\dot{q}}) \\\dot{\bar{k}}(\bm{q},\bm{\dot{q}})
\end{bmatrix} \nonumber \\
&\approx \underbrace{\begin{bmatrix}
\bm{J}_{\textrm{pos}_1}(\bm{q}) + \bar{k}(\bm{q})\bm{J}_{\textrm{FOV}_1}(\bm{q}) + a_{\x}(\bm{q})\bm{J}_{\textrm{dist}_t}(\bm{q}) \\ 
\bm{J}_{\textrm{pos}_2}(\bm{q}) + \bar{k}(\bm{q})\bm{J}_{\textrm{FOV}_2}(\bm{q}) + a_{\y}(\bm{q})\bm{J}_{\textrm{dist}_t}(\bm{q}) \\ \bm{J}_{\textrm{dist}_t}(\bm{q})
\end{bmatrix}}_{\bm{J}_{\textrm{spray}}(\bm{q})}\bm{\dot{q}}.\label{eq:J_spray}
\end{align}

\subsection{IMPLEMENTATION}
\label{sec:implementation}
\begin{figure}[htb]
\centering
\includegraphics[width=0.45\textwidth]{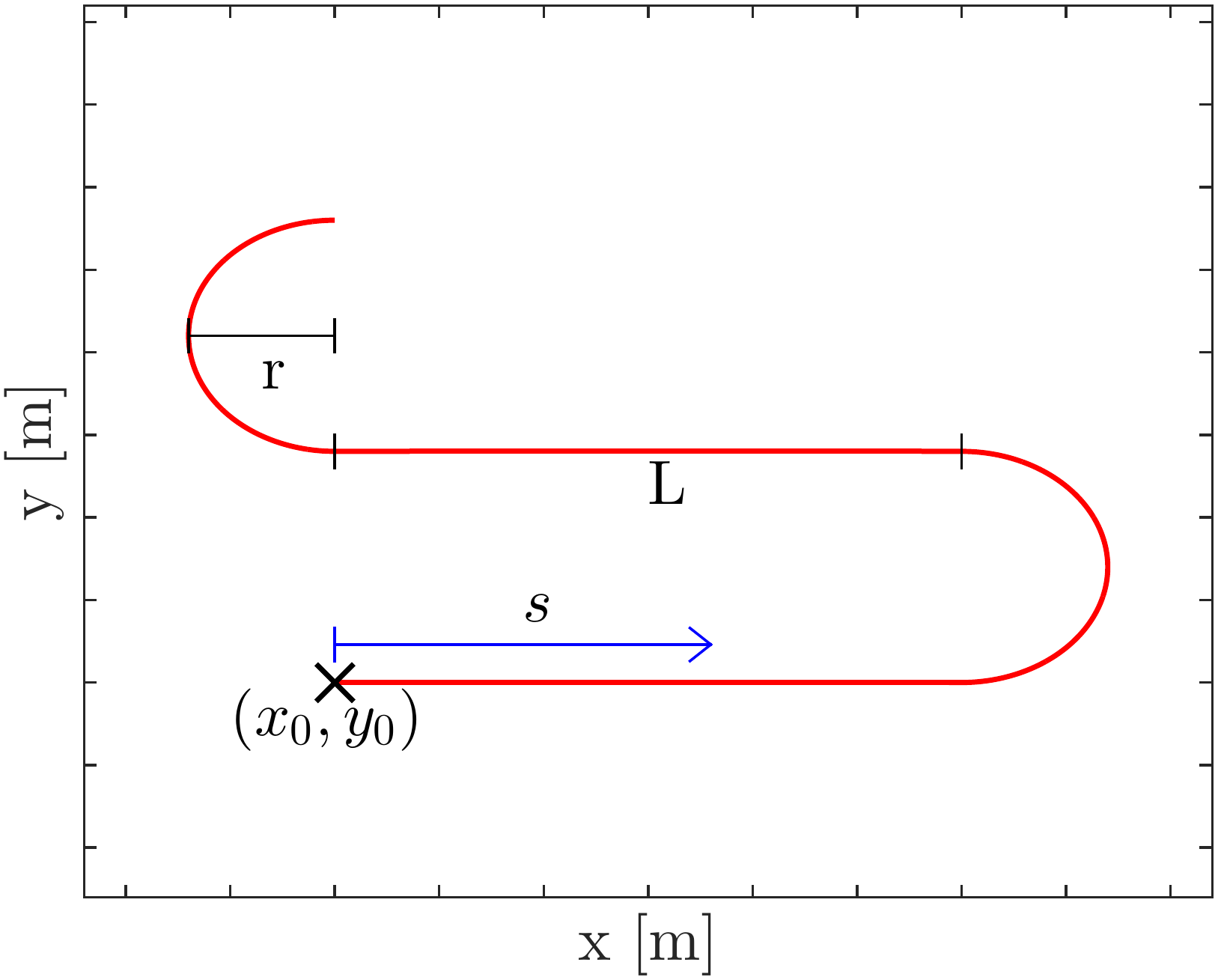}
\caption{Spray pattern fully defined by the length $L$, radius $r$ and initial point $(x_0,y_0)$.}
\label{fig:spraypattern}
\end{figure}

In these experiments, the lawn mowing spray pattern is defined by a length $L$, a radius $r$ and an initial point $(x_0,y_0)$, and is illustrated in Fig.~\ref{fig:spraypattern}. The pattern is parametrized by the arc length $s$:

\footnotesize
\begin{align}
x_{\textrm{spray}}(s) &= \left\lbrace \begin{array}{l}
 s + x_0,~~~~~~~~~~~~~~~~~~~~~~~~~~~~~~~~~~~~~~~   0 \le s \le L \\
 L + r\sin(\frac{s-L}{r}) + x_0,~~~~~~~~~~~~~~~~~~~~~~~~  L <  s  \le L + \pi r \\
 L - (s-L-\pi r) + x_0,~~~~~~~~~~~~~~~~ L + \pi r < s \le 2L + \pi r \\
 -r\sin(\frac{s-2L-\pi r}{r}) + x_0,~~~~~~~~~~~~~~  2L + \pi r < s \le 2(L+\pi r),\end{array} \right. \label{eq:x_spray}\\
y_{\textrm{spray}}(s) &= \left\lbrace \begin{array}{l}
  y_0,~~~~~~~~~~~~~~~~~~~~~~~~~~~~~~~~~~~~~~~~~~~    0 \le  s  \le L \\
  r(-\cos(\frac{s-L}{r})+1)  + y_0,~~~~~~~~~~~~~~~~~~~L <  s  \le L + \pi r \\
  2r + y_0,~~~~~~~~~~~~~~~~~~~~~~~~~~~~~~~   L + \pi r <  s  \le 2L + \pi r \\
  2r+r(\cos(\frac{s-2L-\pi r}{r})-1) + y_0,~~~2L + \pi r <  s  \le 2(L+\pi r),\end{array} \right. \label{eq:y_spray}\\
  &\Downarrow \nonumber \\
 \dot{x}_{\textrm{spray}}(s,\dot{s}) &= \left\lbrace \begin{array}{l}
   \dot{s},~~~~~~~~~~~~~~~~~~~~~~~~~~~~~~~~~~~~~~~~~~~~~    0 \le  s  \le L \\
   \dot{s}\cos(\frac{s-L}{r}),~~~~~~~~~~~~~~~~~~~~~~~~~~~~~~~~~~    L <  s  \le L + \pi r \\
   -\dot{s},~~~~~~~~~~~~~~~~~~~~~~~~~~~~~~~~~~~~~   L + \pi r <  s  \le 2L + \pi r \\
   -\dot{s}\cos(\frac{s-2L-\pi r}{r})~~~~~~~~~~~~~~~~~~~~~   2L + \pi r <  s  \le 2(L+\pi r),\end{array} \right. \\
  \dot{y}_{\textrm{spray}}(s,\dot{s}) &= \left\lbrace \begin{array}{l}
    0,~~~~~~~~~~~~~~~~~~~~~~~~~~~~~~~~~~~~~~~~~~~~~     0 \le  s  \le L \\
    \dot{s}\sin(\frac{s-L}{r}),~~~~~~~~~~~~~~~~~~~~~~~~~~~~~~~~~~~    L <  s  \le L + \pi r \\
    0,~~~~~~~~~~~~~~~~~~~~~~~~~~~~~~~~~~~~~~~   L + \pi r <  s  \le 2L + \pi r \\
    -\dot{s}\sin(\frac{s-2L-\pi r}{r})~~~~~~~~~~~~~~~~~~~~~~   2L + \pi r <  s  \le 2(L+\pi r).\end{array} \right.\label{eq:y_spray_dot}
\end{align}
\normalsize

\noindent Note that the arc length $s$ is a virtual parameter that we are free to choose. In these experiments, we have chosen
\begin{equation}
s(t) = Ut,
\end{equation}
where $U > 0$ is a constant and represents the spray velocity along the surface. Therefore, the spray pattern is in effect a function of time, $x_{\textrm{spray}}(s(t))$ and $y_{\textrm{spray}}(s(t))$. Note that any spray pattern given as a function of time may be applied for the method proposed in this paper, and that the typical spray pattern~(\ref{eq:x_spray})-(\ref{eq:y_spray_dot}) is chosen as an example. Also note that $\dot{s}(t) = U$, and that $U$ is the velocity of the trajectory at the paint surface and not of the end effector itself.

\begin{figure}
\centering
\includegraphics[width=0.7\linewidth]{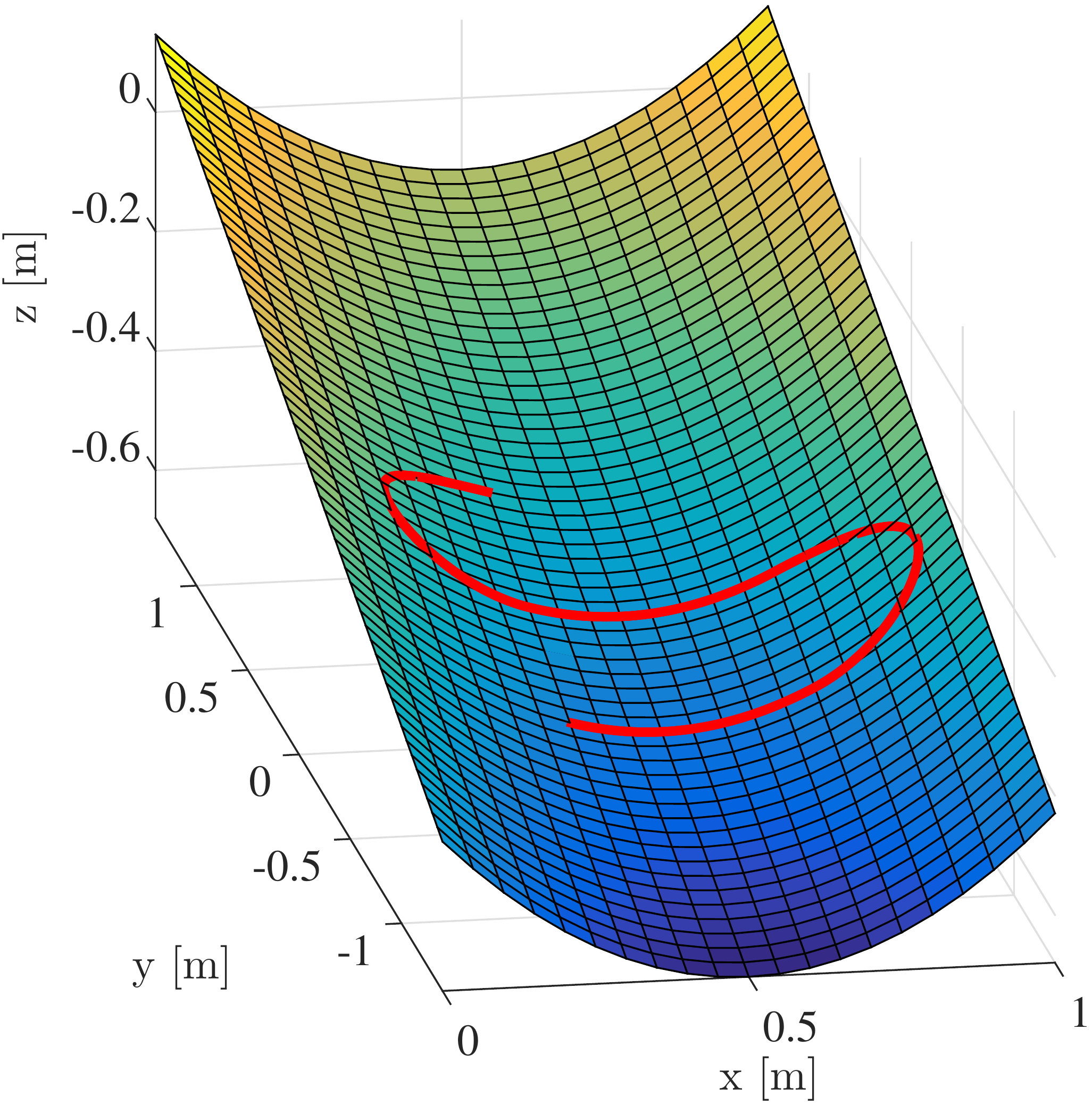}
\caption{Example of spray surface and spray pattern.}
\label{fig:surface_example}
\end{figure}

Furthermore, the surface is described by the function $z = h(x,y)$. For instance, a flat surface on the $xy$-plane would yield $h(x,y) = c$ for some constant $c$, whereas the curved surface given by 
\begin{equation}
z = h(x,y) = (x-0.5)^2+0.2y - 0.4 \label{eq:surface_example}
\end{equation}
and the resulting spray pattern for $L = 0.4$ m, $r = 0.2$ m, $x_0 = 0.35$ m and $y_0 = -0.55$ m are illustrated in Fig.~\ref{fig:surface_example} as an example. The spray pattern is given in spacial coordinates by
\begin{equation}
\bm{v}(s) = \begin{bmatrix} x_{\textrm{spray}}(s)\\y_{\textrm{spray}}(s)\\h(x_{\textrm{spray}}(s),y_{\textrm{spray}}(s))\end{bmatrix}_.
\end{equation}

Having established the spray task, we use FOV as a set-based task to ensure that the angle between the spray direction and the surface normal does not exceed a maximum limit of $\theta$. Hence, according to set-based theory~\cite{Moe2016}, the resulting system has two modes, one where the orientation evolves freely, and one where the FOV error is frozen at the allowed maximum limit. For the former, we denote the system task as
\begin{equation}
\bm{\sigma}_1(\bm{q}) = \bm{\sigma}_{\textrm{spray}}(\bm{q}), \label{eq:sigma_1}
\end{equation}
which is defined in~(\ref{eq:sigma_spray}) and 
\begin{equation}
\bm{\sigma}_{1,\des}(t) = \begin{bmatrix}
x_{\textrm{spray}}(s(t)) \\ y_{\textrm{spray}}(s(t)) \\ \bar{k}_{\des}
\end{bmatrix}_,~~~\bm{\dot{\sigma}}_{1,\des}(t) = \begin{bmatrix}
\dot{x}_{\textrm{spray}}(s(t),\dot{s}(t)) \\ \dot{y}_{\textrm{spray}}(s(t),\dot{s}(t)) \\ 0
\end{bmatrix}_,
\label{eq:sigma_1_des}\end{equation}
where $\bar{k}_{\des}>0$ is a constant desired distance between the end effector and the spray surface along the line $l$ and $x_{\textrm{spray}}(s(t))$, $y_{\textrm{spray}}(s(t))$ and their respective time derivatives are given in~(\ref{eq:x_spray})-(\ref{eq:y_spray_dot}) with $s(t) = Ut$ and $\dot{s}(t) = U$. 

Hence, we define mode 1 as
\begin{equation}
\bm{f}_1(t,\bm{q}) = \bm{J}_{\textrm{spray}}^{\dag}(\bm{q})\left(\bm{\dot{\sigma}}_{1,\des}(t)+\bm{\Lambda}_1\bm{\tilde{\sigma}}_1(t,\bm{q})\right),\label{eq:f1}
\end{equation}
where $\bm{\Lambda}_1 > 0$ is a positive definite matrix and $\tilde{\bm{\sigma}}_1=\bm{\sigma}_{1,\des}-\bm{\sigma}_1$. Here, we control the spray trajectory on the surface (the point where the spray hits the surface) and the distance between the nozzle and the surface. In this mode there is no limitation on the orientation, which could exceed the maximum angle between the surface normal and spray direction, above which the quality of the paint job may deteriorate. Hence, we allow $\bm{\dot{q}}$ to follow the vector field $\bm{f}_1$ as long as this mode will not result in the set-based task being violated, and switch to mode 2 otherwise. Switching is determined by the tangent cone~(\ref{eq:tangent_cone_def}). In mode 2, the FOV-vector is frozen on this limit. Hence, we define
\begin{equation}
\bm{\sigma}_2(\bm{q}) = \begin{bmatrix}
\bm{\sigma}_1(\bm{q}) \\
\sigma_{\textrm{FOV}}(\bm{q})
\end{bmatrix}_,\label{eq:sigma_2}
\end{equation}
where $\sigma_{\textrm{FOV}}(\bm{q})$ is defined in~(\ref{eq:sigma_FOV}) with
\begin{equation}
\bm{a}_{\des}(\bm{q}) = -\frac{N(\bm{q})}{|N(\bm{q})|},
\end{equation}
\begin{equation}
\bm{\sigma}_{2,\des}(t) = \begin{bmatrix}
\bm{\sigma}_{1,\des}(t) \\ \sqrt{2(1-\cos(\theta))}
\end{bmatrix}_,~~~\bm{\dot{\sigma}}_{2,\des}(t) = \begin{bmatrix}\bm{\dot{\sigma}}_{1,\des}(t) \\ 0\end{bmatrix}_,
\label{eq:sigma_2_des}
\end{equation}
where $\theta$ is the maximum allowed degrees between the FOV vector and the surface normal. Hence, we define mode 2 as
\begin{equation}
\bm{f}_2(t,\bm{q}) = \begin{bmatrix}\bm{J}_{\textrm{spray}}(\bm{q}) \\ \bm{J}_{\textrm{FOV}}(\bm{q})\end{bmatrix}^{\dag}\left(\bm{\dot{\sigma}}_{2,\des}+\bm{\Lambda}_2\bm{\tilde{\sigma}}_2\right).\label{eq:f2}
\end{equation}

\noindent The step-by-step implementation for an iteration in the control system is the given below:
\begin{algorithm}
\KwIn{Time $t$, current configuration $\bm{q}$}
Update the variable $s = U(t-t_0)$\;
Calculate $\bm{\sigma}_{1,\des}(t)$, $\bm{\sigma}_{2,\des}(t)$, $\bm{\dot{\sigma}}_{1,\des}(t)$ and $\bm{\dot{\sigma}}_{2,\des}(t)$, eq.~(\ref{eq:sigma_1_des}) and~(\ref{eq:sigma_2_des})\;
Use the forward kinematics to find $\bm{p}_{\e}(\bm{q})$ and $\bm{a}(\bm{q})$\;
Use $\bm{p}_{\e}(\bm{q})$, $\bm{a}(\bm{q})$ and optimization techniques to calculate $\bar{k}(\bm{q})$, eq.~(\ref{eq:bar_k})\;
Calculate $\bm{\sigma}_1(\bm{q})$, $\bm{\sigma}_2(\bm{q})$, $\bm{J}_{\textrm{spray}}(\bm{q})$ and $\bm{J}_{\textrm{FOV}}(\bm{q})$, eq~(\ref{eq:sigma_1}),~(\ref{eq:sigma_2}),~(\ref{eq:J_spray}) and~(\ref{eq:J_FOV})\;
Find $\bm{f}_1(t,\bm{q})$ and $\bm{f}_2(t,\bm{q})$, eq.~(\ref{eq:f1}) and~(\ref{eq:f2})\;
Use the tangent cone function in Alg.~\ref{alg:in_t_c} to decide mode: $$a = \textrm{in\_T\_C}(\bm{J}_{\textrm{FOV}}\bm{f}_1,\sigma_{\textrm{FOV}},0,\sqrt{2(1-\cos(\theta))})$$
\uIf{$a$ is True}{$\bm{\dot{q}}_{\des} = \bm{f}_1$\;}
\Else{$\bm{\dot{q}}_{\des} = \bm{f}_2$\;}
\caption{One iteration in the control system.}
\label{alg:main}
\end{algorithm}

\section{EXPERIMENTAL RESULTS}
\label{sec:experimental_results}
In the experiments, the following parameters have been used:
\begin{align}\\
\theta &= 20^{\circ} \\
h(x,y) &= -0.45 \\
\bar{k}_{\des} &= 0.3 \text{m} \\
\bm{\Lambda}_1 &= 0.4\bm{I}_{3\times 3} \\
\bm{\Lambda}_2 &= 0.4\bm{I}_{4\times 4}
\end{align}
The choice of $\theta = 20^{\circ}$ is the same as in~\cite{From2011}. Since the function $h(x,y)$ is a constant, the spray surface is flat, and the desired distance between the nozzle and the point of intersection $\bm{p}_{\i}$ is 0.3 m. Note that in these experiments, no paint has actually been applied to a surface. The algorithm controls the movement of the manipulator in such a way that it corresponds to a paint scenario. 

\begin{figure}[htbp]
\centering
\begin{tabular}{c}
\begin{subfigure}[h]{0.3\textwidth}
                \centering
                \includegraphics[width=\textwidth]{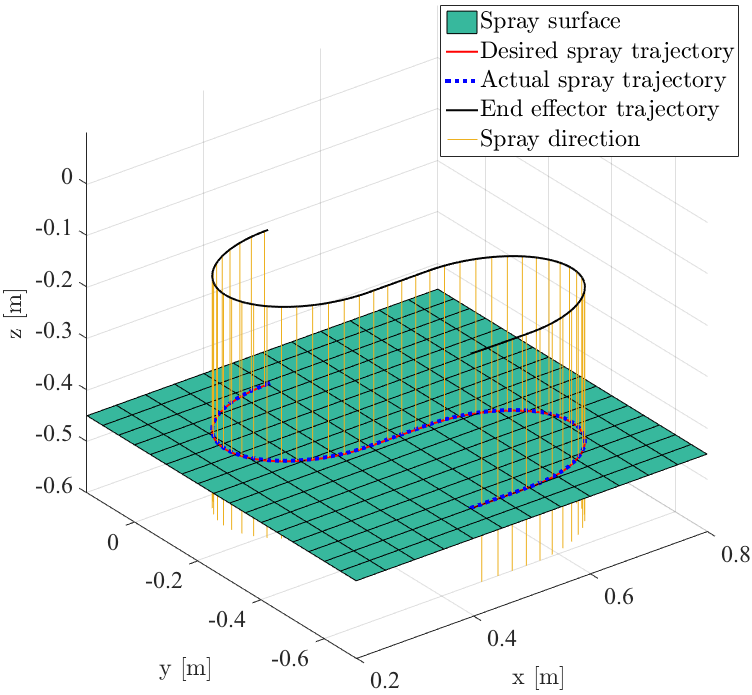}
\end{subfigure}
\\
\begin{subfigure}[h]{0.3\textwidth}
                \centering
                \includegraphics[width=\textwidth]{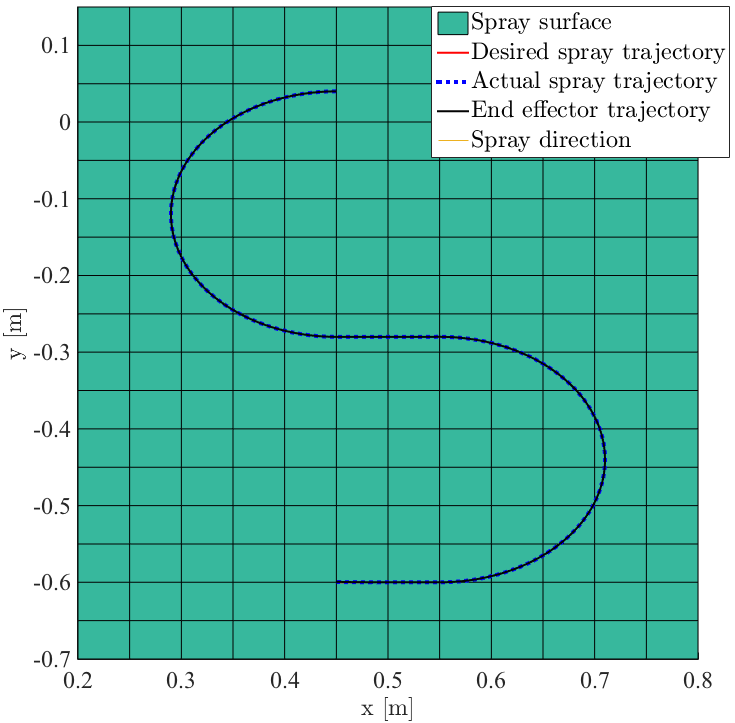}
\end{subfigure}
\\
\begin{subfigure}[h]{0.3\textwidth}
                \centering
                \includegraphics[width=\textwidth]{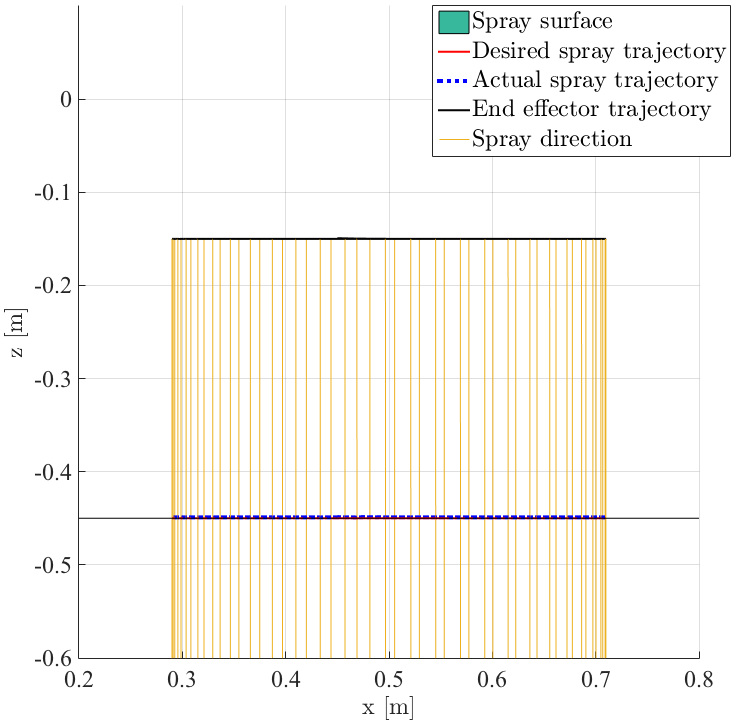}
\end{subfigure}
\end{tabular}
\caption{Current standard solution for spray paint. The paint is applied orthogonally to the spray surface. For a flat surface, the end effector movement copies the pattern at a constant distance.}
\label{fig:standard}
\end{figure}

\begin{figure}[htbp]
\centering
\begin{tabular}{c}
\begin{subfigure}[h]{0.3\textwidth}
                \centering
                \includegraphics[width=\textwidth]{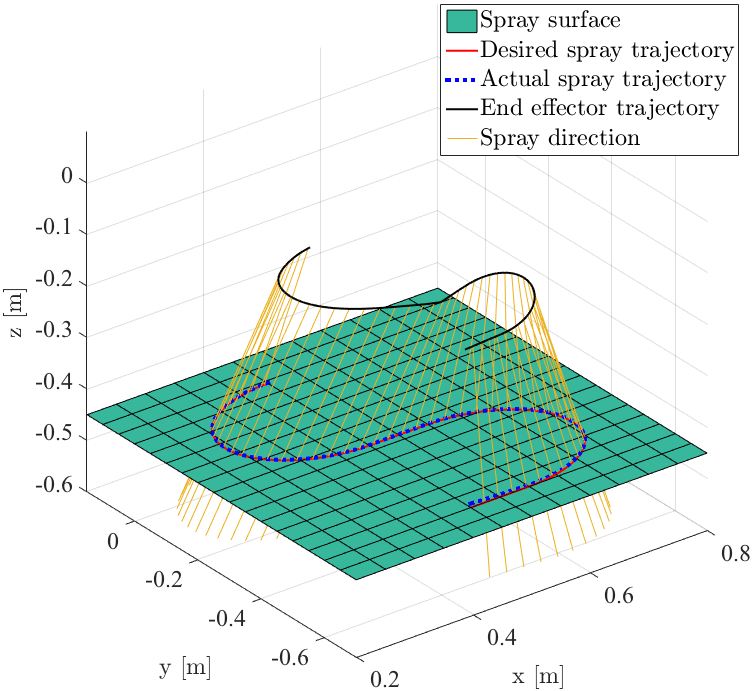}
\end{subfigure}
\\
\begin{subfigure}[h]{0.3\textwidth}
                \centering
                \includegraphics[width=\textwidth]{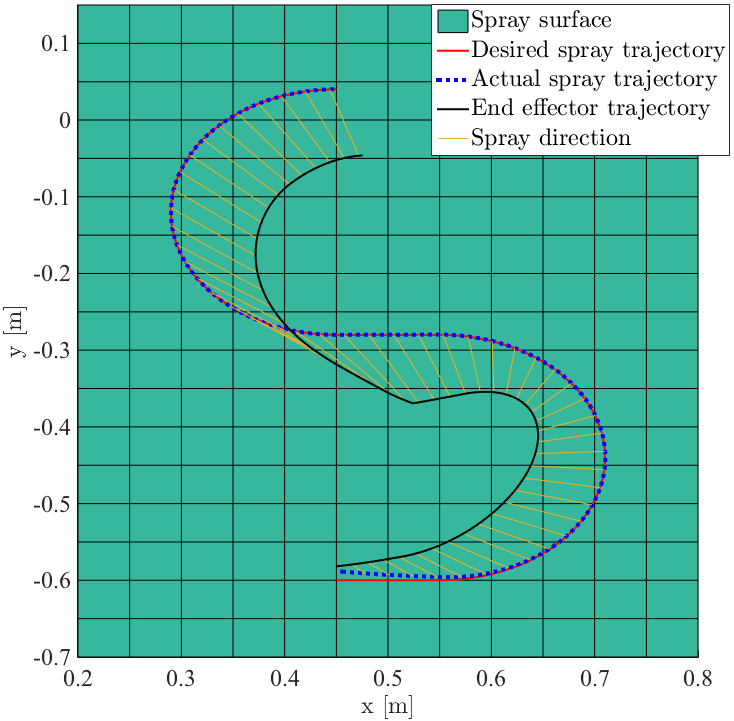}
\end{subfigure}
\\
\begin{subfigure}[h]{0.3\textwidth}
                \centering
                \includegraphics[width=\textwidth]{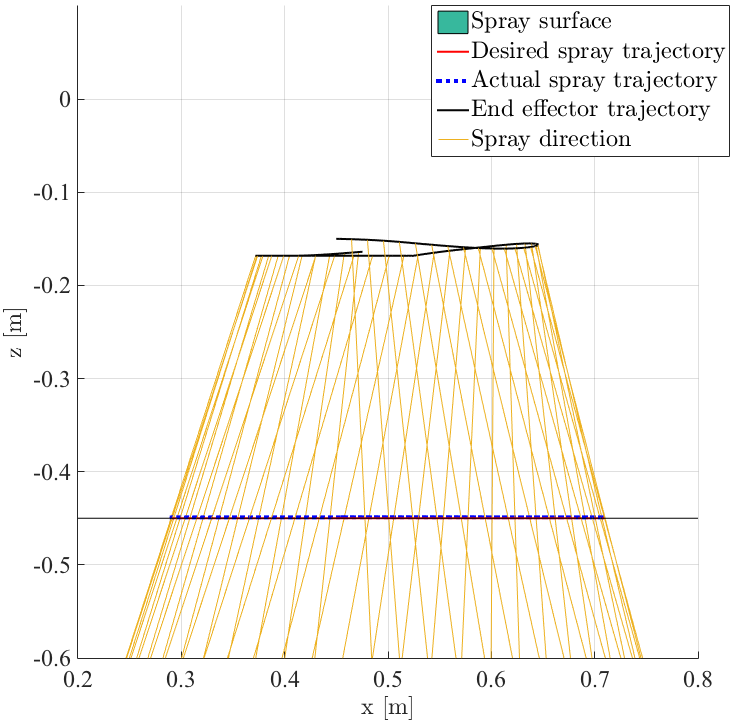}
\end{subfigure}
\end{tabular}
\caption{Set-based solution for spray paint, approach A. The maximum allowed angle between the spray direction and the surface normal is limited to $\theta = 20^{\circ}$. The orientation evolves freely inside this set.}
\label{fig:setbased}
\end{figure}

Experiments have been run for three different patterns and three spray velocities:
\begin{align}
(r,L) &= \left\lbrace(0.07, 0.3),(0.12, 0.2),(0.16,0.1)\right\rbrace \text{m} \\
U &= \left\lbrace 0.15, 0.10, 0.05 \right\rbrace \text{m/s}
\end{align}
Furthermore, several set-based approaches have been tested and compared to the current industry standard, in which the spray gun is kept perpendicular to the spray surface at all times. This corresponds to $\bm{\dot{q}}_{\des} \equiv \bm{f}_2$, with $\theta = 0$. This method is referred to as ST in the tables below. The following set-based approaches have been implemented:
\begin{itemize}
\item A - Pure set-based like described in the step-by-step implementation above
\item B - Switch between ST on the straight segments and approach A on the turns
\item C - Corresponds to A with smooth switching, i.e. when switching between mode 1 and 2
\item D - Corresponds to B with smooth switching
\end{itemize}
To illustrate, the results are plotted below in 3D with the current standard solution ST in Fig.~\ref{fig:standard} and the set-based approach A in Fig.~\ref{fig:setbased}, for $r=0.16$ m, $L = 0.10$ m and $U = 0.10$ m/s.

\begin{figure}[htbp]
\centering
\includegraphics[width=0.5\textwidth]{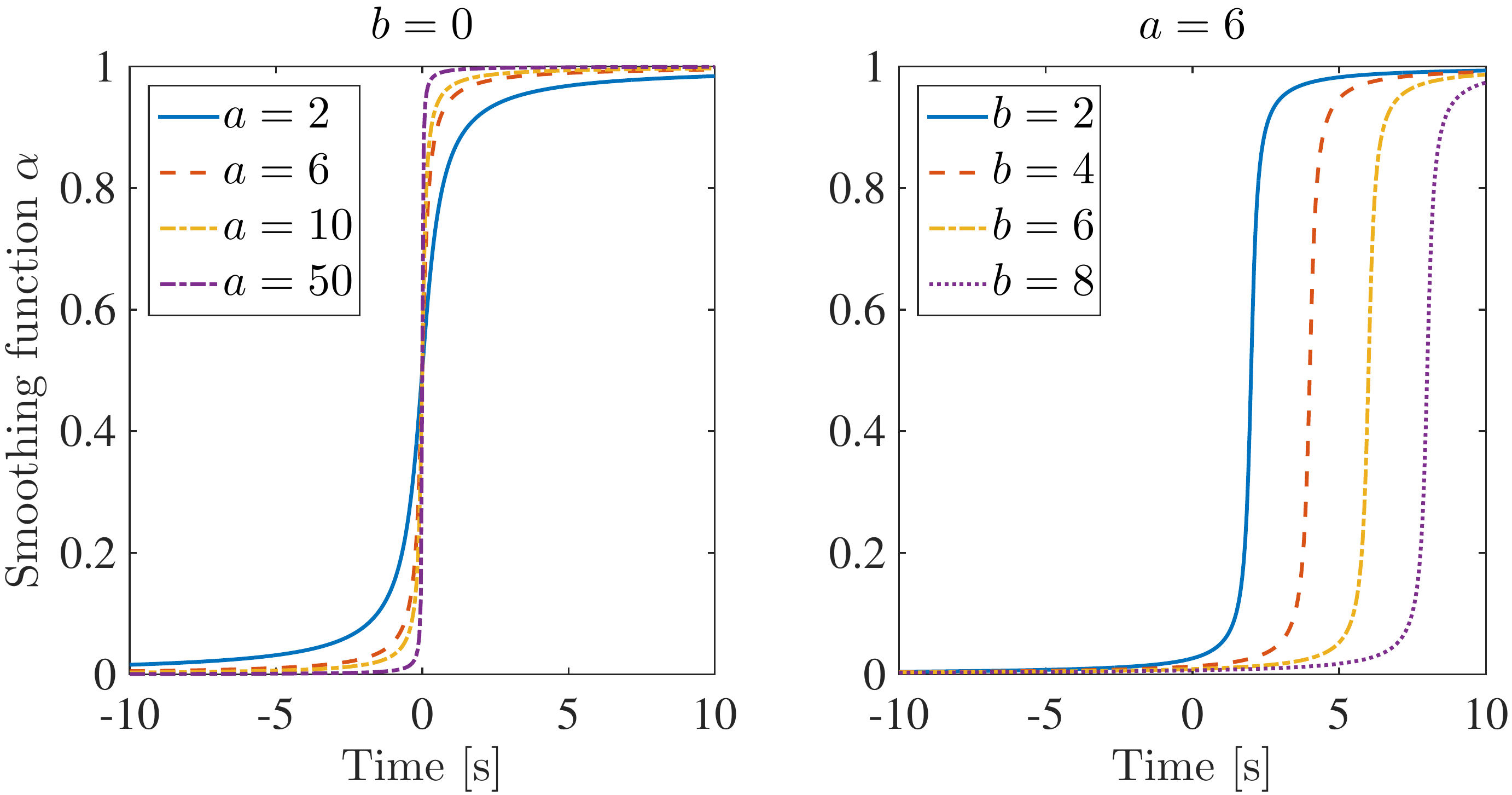}
\caption{Smoothing function illustrated for different parameters and $t_{\text{last\_switch}}=0$. The parameter $a$ determines the sharpness, i.e. how long the transition takes, and $b$ the temporal shift, i.e. how long after the last switch the smoothing function is activated.}
\label{fig:smoothing_function}
\end{figure}

When switching between modes in approach A and B, the reference $\bm{\dot{q}}_{\des}$ changes abruptly, thereby requiring large joint accelerations. This could potentially be very demanding for the manipulator to handle. Hence, in approach C and D switches are detected between mode 1 and 2 and between ST, mode 1 and mode 2, respectively. The reference velocity $\bm{\dot{q}}_{\des}$ is then smoothened between the previous and the new reference using the following smoothing function:
\begin{equation}
\alpha(t,t_{\text{last\_switch}}) = \frac{1}{\pi}\arctan\left(a(t-t_{\text{last\_switch}}-b)\right)+\frac{1}{2}
\label{eq:alpha}
\end{equation}
Here, the parameters $a$ and $b$ determine the sharpness of smoothing function and the delay in time before the transition takes place, respectively. This is illustrated in Fig.~\ref{fig:smoothing_function}. In these experiments, $a = 110$ and $b = 0.05$.

Note that when smooth switching is used, the strict priority of the modes are lost during transition. In mode 1, the FOV task evolves freely, and in mode 2 the task is kept stationary at the maximum limit of $\theta$. Hence, if mode 2 is activated at the maximum limit and smooth switching is applied to the transition, this limit might be exceeded because mode 1 affects the solution $\bm{\dot{q}}_{\des}$ during the transition. To avoid exceeding the limit, step 7 of the implementation in Alg.~\ref{alg:main} is slightly different for approaches C and D. In this case, the boolean variable $a$ is defined as
\begin{equation}
a = \textrm{in\_T\_C}(\bm{J}_{\textrm{FOV}}\bm{f}_1,\sigma_{\textrm{FOV}},0,\sqrt{2(1-\cos(\theta-\theta_0))}),
\end{equation}
where $\theta_0$ has been chosen as $5^{\circ}$. For practical purposes, this means that if the FOV angle between the spray direction and spray surface exceeds $15^{\circ}$, mode 2 is activated and this angle is slowly controlled towards the maximum allowed limit of $\theta=20^\circ$. This approach is conservative, but by defining this buffer zone of $\theta_0$, we ensure that the maximum limit is not exceeded when smooth switching is applied.

\begin{figure}[htbp]
	\centering
    \includegraphics[width=0.5\textwidth]{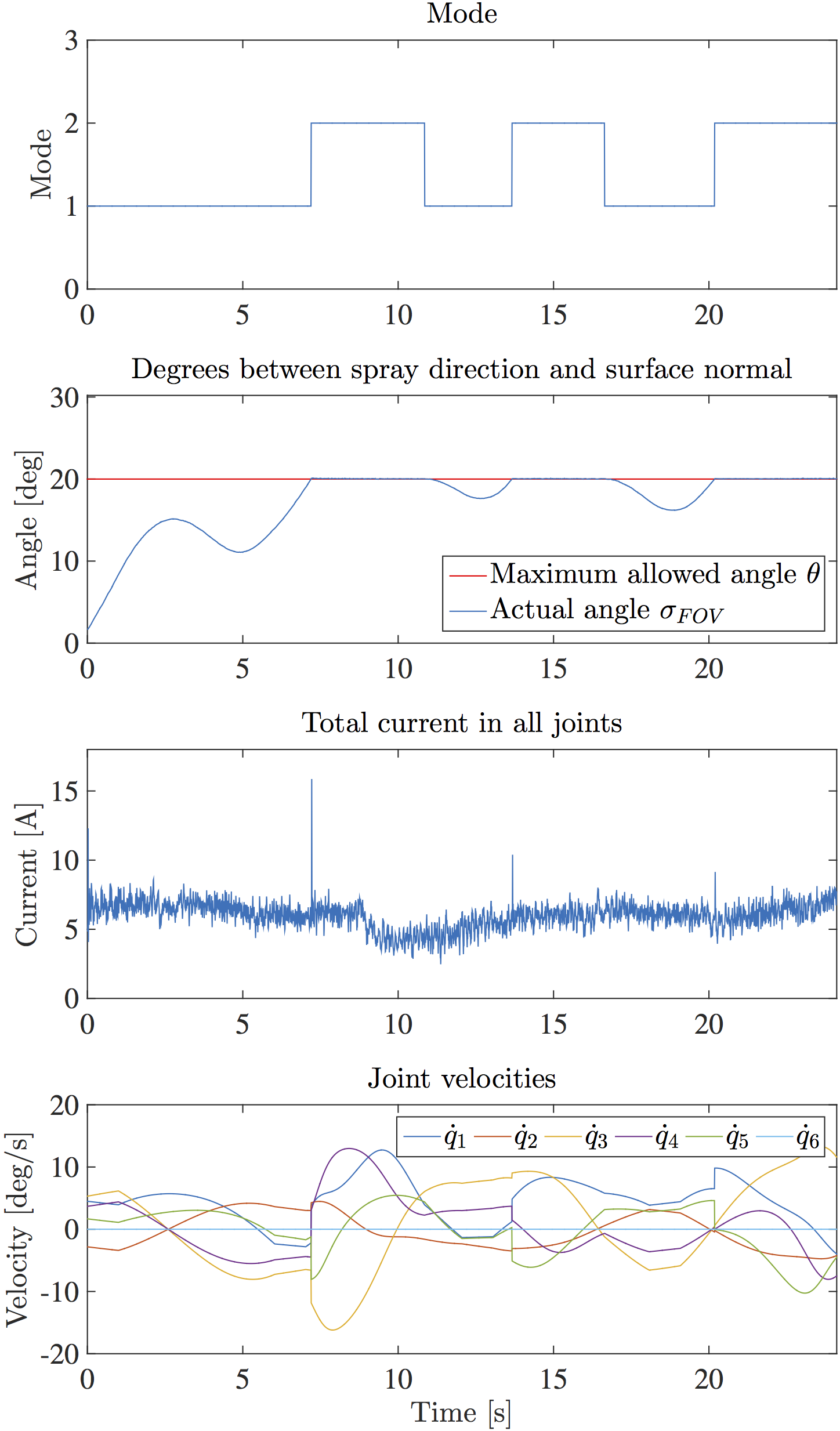}
	\caption{Approach A, $U = 0.10$ m/s, $r=0.16$ m, $L = 0.10$. Active mode over time, set-based task $\sigma_{\textrm{FOV}}(\bm{q})$, total current in all joints and joint velocities. In this approach, the set-based task $\sigma_{\textrm{FOV}}$ evolves freely until the maximum limit $\theta$, but the switch between modes results in spikes in current and large joint accelerations.\vspace{\baselineskip}}
	\label{fig:a6}
\end{figure}

\begin{figure}[htbp]
	\centering
    \includegraphics[width=0.4982\textwidth]{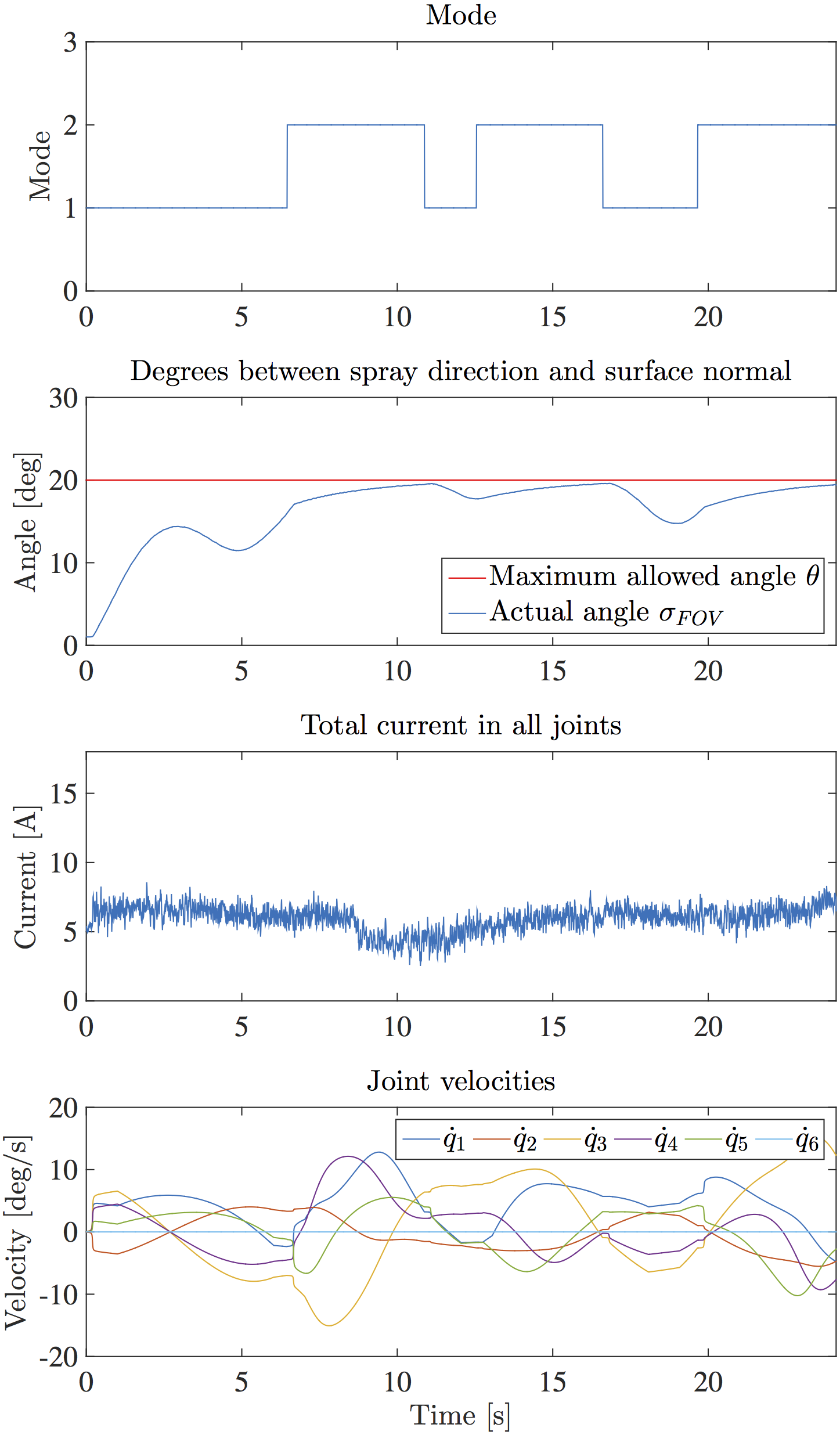}
	\caption{Approach C, $U = 0.10$ m/s, $r=0.16$ m, $L = 0.10$. Active mode over time, set-based task $\sigma_{\textrm{FOV}}(\bm{q})$, total current in all joints and joint velocities. In this approach, the set-based task $\sigma_{\textrm{FOV}}$ evolves freely until a buffer limit $(\theta-\theta_0)$, after which it is smoothly controlled towards the maximum limit. In this more conservative approach there are no large peaks in the current and the joint velocities are smooth.}
	\label{fig:c6}
\end{figure}

To calculate energy consumption, we measure the total current in all the joints, see Fig.~\ref{fig:a6} and~\ref{fig:c6}. The UR5 runs on a 24 V supply supply\footnote{UR5 user manual: \url{https://www.universal-robots.com/media/8704/ur5_user_manual_gb.pdf}}, and the mean power is therefore given as
\begin{equation}
\bar{P} = 24\bar{I}_{\textrm{tot}},
\end{equation}
where $\bar{I}_{\textrm{tot}}$ is the mean of the current in all the joints. The total energy consumption is then given as
\begin{equation}
E = \bar{P}\Delta t,
\end{equation}
where $\Delta t$ is the time of the experiment.

The experimental results are shown in Table~\ref{tab:experimental_results_mean_vel} and~\ref{tab:experimental_results_max_vel}. In Table~\ref{tab:experimental_results_mean_vel}, the standard approach is compared to the set-based approached A-D for the same spray velocity $U$. In all experiments, the spray pattern has been repeated twice, and as the Table~\ref{tab:experimental_results_mean_vel} shows, for the standard solution the length of the end effector path is equal to the spray pattern length $2(2\pi r + 2L)$, and the average end effector velocity is equal to the spray velocity $U$. This is expected, since the end effector in this approach simply copies the spray pattern while keeping the angle between the spray direction and surface equal to zero (see Fig.~\ref{fig:standard}). However, for the set-based approaches, by allowing an angle between the spray direction and the surface normal, the end effector moves a shorter distance while maintaining the same spray velocity $U$ at the spray surface. Therefore, the average velocity of the end effector is lower than $U$. The set-based approaches A-D thus consume less energy that the standard method ST of controlling the orientation directly, on average 4.32\% less, and at best, 8.05\%. Since the spray velocity $U$ is the same for the set-based and standard solutions, the approaches spend the same amount of time to complete the spray pattern. Furthermore, the maximum angle $\theta = 20^{\circ}$ between the spray direction and the surface normal is never exceeded.

\begin{table*}[htbp]
\begin{tabular}{|p{1.1cm} | p{2cm}  | p{0.5cm} p{0.5cm} p{0.5cm} p{0.5cm} p{0.5cm}  | p{0.5cm} p{0.5cm} p{0.5cm} p{0.5cm} p{0.5cm} | p{0.5cm} p{0.5cm} p{0.5cm} p{0.5cm} p{0.5cm}|}
\hline
\multicolumn{2}{|c|}{} & \multicolumn{5}{|c|}{U = 0.15 m/s} & \multicolumn{5}{|c|}{U = 0.10 m/s} & \multicolumn{5}{|c|}{U = 0.05 m/s} \\
 \cline{3-17}
\multicolumn{2}{|c|}{} & ST & A & B & C & D & ST & A & B & C & D & ST & A & B & C & D \\
 \hline
\multirow{4}{1.3cm}{r = 0.07 m  L = 0.3 m} & Time [s] & 13,87 & 13,87 & 13,87 & 13,86 & 13,86 & 20,80 & 20,80 & 20,80 & 20,80 & 20,80 & 41,60 & 41,60 & 41,60 & 41,60 & 41,60\\ & Ee path [m] & 2,08	& 1,39	& 1,69	& 1,43& 1,75 & 2,08	& 1,39& 1,73& 1,42 & 1,78 & 2,08	& 1,39	& 1,79	& 1,41	& 1,82\\
& Avg. ee vel. [m/s] & 0,15	& 0,10	& 0,12	& 0,10	& 0,13 & 0,10	& 0,07	& 0,08	& 0,07	& 0,09 & 0,05	& 0,03	& 0,04	& 0,03	& 0,04\\
& Energy [\%] & 100 &	95,67 &	98,70 &	95,27 &	97,94 & 100 & 97,35 & 99,99 & 96,63 &	97,93 & 100 & 97,11 & 98,99 & 97,30 & 99,86 \\
\hline
\multirow{4}{1.3cm}{r = 0.12 m  L = 0.2 m} & Time [s] & 15,39 & 15,39 & 15,39 & 15,39 & 15,39 & 23,08 & 23,09 & 23,09 & 23,08 & 23,08 & 46,17 & 46,17 & 46,16 & 46,16 & 46,16 \\ & Ee path [m] & 2,31	& 1,63	& 1,70	& 1,62	& 1,73 & 2,31 & 1,62 & 1,69 & 1,62 & 1,74 & 2,31 & 1,62 & 1,78 & 1,62 & 1,81\\
& Avg. ee vel. [m/s] & 0,15 & 0,11 & 0,11 & 0,11 & 0,11 & 0,10 & 0,07 & 0,07 & 0,07 & 0,08 & 0,05 & 0,04 & 0,04 & 0,04 & 0,04 \\
& Energy [\%] & 100 & 93,25 & 96,99 & 92,62 & 95,37 & 100 & 92,50 &	94,82 &	91,95 &	93,71 & 100 & 94,83 & 96,77 & 94,17 & 97,34 \\
\hline
\multirow{4}{1.3cm}{r = 0.16 m L = 0.1 m} & Time [s] & 16,07 & 16,07 & 16,07 & 16,07 & 16,07 & 24,11 & 24,11 & 24,11 & 24,11 & 24,11 & 48,22 & 48,22 & 48,22 & 48,21 & 48,21 \\ & Ee path [m] & 2,42 & 1,78 & 1,82 & 1,77 & 1,83 & 2,42 & 1,78 & 1,82 & 1,78 & 1,83 & 2,42 & 1,78 & 1,81 & 1,77 & 1,83\\
& Avg. ee vel. [m/s] & 0,15	& 0,11 & 0,11 & 0,11 & 0,11 & 0,10 & 0,07 & 0,08 & 0,07 & 0,08 & 0,05 & 0,04 & 0,04 & 0,04 & 0,04\\
& Energy [\%] & 100 & 93,84 & 95,47 & 93,11 & 94,35 & 100 & 95,08 &	96,24 &	94,30 &	95,61 & 100 & 93,91 & 95,90 & 94,06 & 95,56 \\
\hline
\end{tabular}
\caption{Experimental results for three spray velocities $U$ and spray patterns $(r,L)$. In each run, the standard method ST is compared to 4 set-based approaches A-D in terms of time to complete spray task, length of end effector path, average end effector velocity and energy consumption.}
\label{tab:experimental_results_mean_vel}
\end{table*}

\begin{table*}[htbp]
\begin{tabular}{|p{1.1cm} | p{2cm}  | p{0.5cm} p{0.5cm} p{0.5cm} p{0.5cm} p{0.5cm}  | p{0.5cm} p{0.5cm} p{0.5cm} p{0.5cm} p{0.5cm} | p{0.5cm} p{0.5cm} p{0.5cm} p{0.5cm} p{0.5cm}|}
\hline
\multicolumn{2}{|c|}{} & \multicolumn{5}{|c|}{End effector avg. vel = 0.15 m/s} & \multicolumn{5}{|c|}{End effector avg. vel = 0.10 m/s} & \multicolumn{5}{|c|}{End effector avg. vel = 0.05 m/s} \\
\cline{3-17}
\multicolumn{2}{|c|}{}  & ST & A & B & C & D & ST & A & B & C & D & ST & A & B & C & D \\
 \hline
\multirow{4}{1.3cm}{r = 0.07 m  L = 0.3 m} & Time [s] & 13,87 & 9,28 & 11,28 & 9,51 & 11,64 & 20,80 & 13,88 & 17,34 & 14,15 & 17,79 & 41,60 & 27,77 & 35,83 & 28,23 & 36,39\\ & Ee path [m] & 2,08	& 1,39 & 1,67 & 1,40 & 1,73 & 2,08 & 1,39 & 1,72 & 1,40 & 1,77 & 2,08 & 1,39 & 1,78 & 1,42 & 1,81 \\
& U [m/s] & 0,15	& 0,22 & 0,18 & 0,22 & 0,18 & 0,10	& 0,15 & 0,12 & 0,15 & 0,12 & 0,05	& 0,08 & 0,06 & 0,07 & 0,06\\
& Energy [\%] & 100 & 66,37 & 82,07 & 67,18 & 83,17 & 100 & 65,02 &	82,99 &	66,41 &	84,47 & 100 & 65,62	& 85,62 & 66,67 & 87,03 \\
\hline
\multirow{4}{1.3cm}{r = 0.12 m  L = 0.2 m} & Time [s] & 15,39 & 10,84 & 11,33 & 10,83 & 11,51 & 23,08 & 16,16 & 16,91 & 16,17 & 17,38 & 46,17 & 32,34 & 35,52 & 32,32 & 36,21\\ & Ee path [m] & 2,31 & 1,62 & 1,70 & 1,62 & 1,72 & 2,31 & 1,62 & 1,69 & 1,61 & 1,73 & 2,31 & 1,62 & 1,74 & 1,62 & 1,78\\
& U [m/s] & 0,15 & 0,21 & 0,20 & 0,21 & 0,20 & 0,10 & 0,14 & 0,14 & 0,14 & 0,13 & 0,05 & 0,07 & 0,07 & 0,07 & 0,06 \\
& Energy [\%] & 100 & 67,67 & 73,71 & 67,63 & 73,22 & 100 & 65,84 &	70,98 &	65,48 &	71,90 & 100 & 66,70 & 75,52 & 66,92 & 76,16 \\
\hline
\multirow{4}{1.3cm}{r = 0.16 m L = 0.1 m} & Time [s] & 16,07 & 11,85 & 12,14 & 11,81 & 12,23 & 24,11 & 17,78 & 18,15 & 17,76 & 18,34 & 48,22 & 35,69 & 36,27 & 35,45 & 36,62 \\ & Ee path [m] & 2,42 & 1,76 & 1,83 & 1,77 & 1,84 & 2,42 & 1,78 & 1,82 & 1,78 & 1,83 & 2,42 & 1,77 & 1,81 & 1,77 & 1,83\\
& U [m/s] & 0,15 & 0,20	& 0,20 & 0,20 & 0,20 & 0,10 & 0,14 & 0,13 & 0,14 & 0,13 & 0,05 & 0,07 & 0,07 & 0,07 & 0,07\\
& Energy [\%] & 100 & 73,68 & 73,69 & 70,28 & 72,96 & 100 & 71,28 &	73,32 &	70,54 &	73,45 & 100 & 71,92 & 72,73 & 70,11 & 73,05 \\
\hline
\end{tabular}
\caption{Experimental results for three end effector velocities and spray patterns $(r,L)$. In each run, the standard method ST is compared to 4 set-based approaches A-D in terms of time to complete spray task, length of end effector path, spray velocity $U$ and energy consumption.}
\label{tab:experimental_results_max_vel}
\end{table*}

Maintaining a constant, high speed during turns may be challenging for a robotic system because of the large torques required. In the case that the specifications of the robot is the limiting factor in the spray process a set-based approach may increase the velocity, since the end effector in this case is not required to copy the pattern directly, but can use the freedom in orientation to achieve the same spray result with a less demanding trajectory for the end effector. To investigate the consequences of increasing the end effector speed, the experiments from Table~\ref{tab:experimental_results_mean_vel} were repeated, but in this case $U$ was adjusted so that the average velocity of the end effector was the same in all experiments. The results are shown in Table~\ref{tab:experimental_results_max_vel}. 

By increasing the spray velocity $U$ such that the resulting average end effector velocity is the same for all approaches, the set-based approaches completed the spray task in less time. On average, the approaches A-D spent 74.50\% of the time compared to the standard solution, and at best 66.73\%. The energy consumption is also significantly reduced, and the set-based approaches in this case consume on average 27.46\%, and at best 34.98\%, less than the standard approach. The main reason for this is the reduced operation time, but as observed in Table~\ref{tab:experimental_results_mean_vel}, the set-based approaches are also more energy efficient than the standard solution in general.

In addition, we can conclude that the pure set-based approach A is more energy and time efficient than approach B, which can be seen as a hybrid between set-based (active on the turns) and the standard solution (active on the straight lines). This is also reflected in the smooth switching versions, where C is more energy and time efficient than D. Future work will therefore be based on approach A and C. 

When comparing abrupt and smooth switching, Table~\ref{tab:experimental_results_mean_vel} and~\ref{tab:experimental_results_max_vel} show that the smooth approaches C and D perform similarly to their respective original versions A and B. However, the smooth switching ensures that the commanded joint accelerations are not too high, making these approaches feasible also for higher velocities that would result in a security stop for the abrupt methods A and B. Approach C and D even consume less energy than A and B when the velocities are large, because in this case the abrupt switches require a large amount of energy compared to the more conservative, yet smooth methods. This is confirmed by Fig.~\ref{fig:a6} and~\ref{fig:c6}, where the active mode, set-based task $\sigma_{\textrm{FOV}}(\bm{q})$, total current in all joints and joint velocities are plotted for approach A and C with $U = 0.10$ m/s, $r=16$ m, $L = 0.1$ m. In mode 1, the set-based task evolves freely according to the spray task, and in mode 2 it is controlled to the maximum limit of $\theta^{\circ}$. In approach A, mode 2 is activated immediately when the set-based task reaches this limit, and we see that the resulting energy consumption peaks, and that the joint velocities change abruptly, indicating very large joint accelerations. In approach C, mode 2 is activated at $\sigma_{\textrm{FOV}}(\bm{q}) = \theta - \theta_0 = 15^{\circ}$, and the transition between mode 1 and 2 is smoothened by the function~(\ref{eq:alpha}). This approach is more conservative, but there are no peaks in the energy consumption, and the joint velocities are much smoother.

\section{CONCLUSION}
\label{sec:conclusion}
This paper presents an algorithm to use set-based control in a spray paint scenario of a general surface. We exploit the fact that a small angle between the spray direction and the surface normal does not affect the quality of the paint, and define this angle as a set-based task with a maximum allowed value. An algorithm is developed to generate reference trajectories for a general robotic system that, if tracked, ensure achievement of the spray task and satisfaction of the set-based task. In other words, a desired spray pattern is tracked on the spray surface with some constant distance between the spray nozzle and the surface, and the maximum angle between the spray direction and the surface normal is never exceeded. In this paper, the spray pattern is represented as a mowing-the-lawn pattern, but the proposed method is also applicable to other patterns.

Four different set-based approaches have been implemented and experimentally verified for three different spray velocities and three spray patterns on a flat surface. Two of these approaches include smooth switching between different modes of the system. In doing so, it can no longer be guarantees that the set-based task will always be satisfied, but the smooth switches ensure that the joint accelerations are not too large and are therefore suitable for larger spray velocities. Furthermore, in the experiments, the set-based task was never violated for any of the approaches. 

Compared to the industry standard, the set-based approaches all consume less energy for the same spray velocity $U$. Furthermore, the set-based approaches require less torque in the turns, and is therefore applicable for higher spray velocities than the industry standard. By adjusting the spray velocity $U$ so that the average end effector velocity is the same in all approaches, the set-based methods on average spent 74.50\% of the time and 72.54\% of the total energy compared to the current industry standard. 


\section*{ACKNOWLEDGMENTS}
This work was supported by the Research Council of Norway through the Center of Excellence funding scheme, project number 223254.

\bibliographystyle{IEEEtran}
\bibliography{library}

\begin{thebibliography}{10}
\providecommand{\url}[1]{#1}
\csname url@rmstyle\endcsname
\providecommand{\newblock}{\relax}
\providecommand{\bibinfo}[2]{#2}
\providecommand\BIBentrySTDinterwordspacing{\spaceskip=0pt\relax}
\providecommand\BIBentryALTinterwordstretchfactor{4}
\providecommand\BIBentryALTinterwordspacing{\spaceskip=\fontdimen2\font plus
\BIBentryALTinterwordstretchfactor\fontdimen3\font minus
  \fontdimen4\font\relax}
\providecommand\BIBforeignlanguage[2]{{%
\expandafter\ifx\csname l@#1\endcsname\relax
\typeout{** WARNING: IEEEtran.bst: No hyphenation pattern has been}%
\typeout{** loaded for the language `#1'. Using the pattern for}%
\typeout{** the default language instead.}%
\else
\language=\csname l@#1\endcsname
\fi
#2}}

\bibitem{Caccavale2001}
F.~Caccavale and B.~Siciliano, ``{Kinematic control of redundant free-floating
  robotic systems},'' \emph{Advanced Robotics}, vol.~15, no.~4, pp. 429--448,
  2001.

\bibitem{Egeland1998}
O.~Egeland and K.~Y. Pettersen, ``{Free-floating robotic systems},'' in
  \emph{Control Problems in Robotics and Automation}.\hskip 1em plus 0.5em
  minus 0.4em\relax Springer Berlin Heidelberg, 1998, vol. 230, pp. 119--134.

\bibitem{Nenchev1992}
D.~Nenchev, Y.~Umetani, and {Kazuya Y.}, ``{Analysis of a redundant free-flying
  spacecraft/manipulator system},'' \emph{IEEE Transactions on Robotics and
  Automation}, vol.~8, no.~1, pp. 1--6, 1992.

\bibitem{Buss2004}
\BIBentryALTinterwordspacing
S.~R. Buss, ``{Introduction to inverse kinematics with jacobian transpose,
  pseudoinverse and damped least squares methods},'' Tech. Rep., 2009.
  [Online]. Available:
  \url{https://www.math.ucsd.edu/sbuss/ResearchWeb/ikmethods/iksurvey.pdf}
\BIBentrySTDinterwordspacing

\bibitem{Klein1983}
C.~A. Klein and C.~Huang, ``{Review of pseudoinverse control for use with
  kinematically redundant manipulators},'' \emph{IEEE Transactions on Systems,
  Man, and Cybernetics}, vol. SMC-13, no.~2, pp. 245--250, 1983.

\bibitem{Siciliano2008}
B.~Siciliano and O.~Khatib, Eds., \emph{{Springer Handbook of Robotics}}.\hskip
  1em plus 0.5em minus 0.4em\relax Berlin, Heidelberg: Springer Berlin
  Heidelberg, 2008.

\bibitem{Siciliano2009}
B.~Siciliano, L.~Sciavicco, L.~Villani, and G.~Oriolo, \emph{{Robotics:
  modelling, planning and control}}.\hskip 1em plus 0.5em minus 0.4em\relax
  Springer Verlag, 2009.

\bibitem{Antonelli2008}
G.~Antonelli, F.~Arrichiello, and S.~Chiaverini, ``{The null-space-based
  behavioral control for autonomous robotic systems},'' \emph{Intelligent
  Service Robotics}, vol.~1, no.~1, pp. 27--39, 2008.

\bibitem{Antonelli2009a}
G.~Antonelli, ``{Stability Analysis for Prioritized Closed-Loop Inverse
  Kinematic Algorithms for Redundant Robotic Systems},'' \emph{IEEE
  Transactions on Robotics}, vol.~25, no.~5, pp. 985--994, oct 2009.

\bibitem{Antonelli1998}
G.~Antonelli and S.~Chiaverini, ``{Task-priority redundancy resolution for
  underwater vehicle-manipulator systems},'' in \emph{Proc. IEEE International
  Conference on Robotics and Automation}, vol.~1, no. May, 1998, pp. 768--773.

\bibitem{Maciejewski1985}
A.~A. Maciejewski and C.~A. Klein, ``{Obstacle Avoidance for Kinematically
  Redundant Manipulators in Dynamically Varying Environments},'' \emph{The
  International Journal of Robotics Research}, vol.~4, no.~3, pp. 109--117, sep
  1985.

\bibitem{Shiyou2011}
D.~Shiyou, Z.~Xiaoping, and L.~Guoqing, ``{The Null-Space-Based Behavioral
  Control for Swarm Unmanned Aerial Vehicles},'' in \emph{Proc. 2011 First
  International Conference on Instrumentation, Measurement, Computer,
  Communication and Control}.\hskip 1em plus 0.5em minus 0.4em\relax IEEE, oct
  2011, pp. 1003--1006.

\bibitem{Tevatia2000}
G.~Tevatia and S.~Schaal, ``{Inverse kinematics for humanoid robots},'' in
  \emph{Proc. 2000 ICRA. Millennium Conference. IEEE International Conference
  on Robotics and Automation. Symposia Proceedings}, vol.~1.\hskip 1em plus
  0.5em minus 0.4em\relax IEEE, 2000, pp. 294--299.

\bibitem{Unicas}
Unicas, ``{Modeling and control for the MARIS UVMS}, Tech. Rep.~5.

\bibitem{Marchand1996}
E.~Marchand, F.~Chaumette, and A.~Rizzo, ``{Using the task function approach to
  avoid robot joint limits and kinematic singularities in visual servoing},''
  in \emph{Proc. IEEE/RSJ International Conference on Intelligent Robots and
  Systems}, vol.~3, 1996, pp. 1083--1090.

\bibitem{Hanafusa1981}
H.~Hanafusa, T.~Yoshikawa, and Y.~Nakamura, ``{Analysis and control of
  articulated robot arms with redundancy},'' in \emph{Proc. 8th ZFAC World
  Congress}, 1981.

\bibitem{Antonelli2015}
G.~Antonelli, S.~Moe, and K.~Y. Pettersen, ``{Incorporating Set-based Control
  within the Singularity-robust Multiple Task-priority Inverse Kinematics},''
  in \emph{Proc. 23rd Mediterranean Conference on Control and Automation},
  2015.

\bibitem{Moe2015}
S.~Moe, A.~R. Teel, G.~Antonelli, and K.~Y. Pettersen, ``{Stability Analysis
  for Set-based Control within the Singularity-robust Multiple Task-priority
  Inverse Kinematics Framework},'' in \emph{Proc. 54th IEEE Conference on
  Decision and Control}, Osaka, Japan, 2015.

\bibitem{Moe2015a}
S.~Moe, G.~Antonelli, K.~Y. Pettersen, and J.~Schrimpf, ``{Experimental Results
  for Set-based Control within the Singularity-robust Multiple Task-priority
  Inverse Kinematics Framework},'' in \emph{Proc. IEEE International Conference
  on Robotics and Biomimetics}, 2015.

\bibitem{From2010a}
P.~J. From and J.~T. Gravdahl, ``{A Real-Time Algorithm for Determining the
  Optimal Paint Gun Orientation in Spray Paint Applications},'' \emph{IEEE
  Transactions on Automation Science and Engineering}, vol.~7, no.~4, pp.
  803--816, oct 2010.

\bibitem{Suh1991}
\BIBentryALTinterwordspacing
S.~H. Suh, I.~K. Woo, and S.~K. Noh, ``{Development of an automatic trajectory
  planning system (ATPS) for spray painting robots},'' in \emph{Proc. 1991 IEEE
  International Conference on Robotics and Automation}.\hskip 1em plus 0.5em
  minus 0.4em\relax IEEE Comput. Soc. Press, 1991, pp. 1948--1955. [Online].
  Available: \url{http://ieeexplore.ieee.org/document/131912/}
\BIBentrySTDinterwordspacing

\bibitem{Kim2003a}
T.~Kim and S.~Sarma, ``{Optimal sweeping paths on a 2-manifold: a new class of
  optimization problems defined by path structures},'' \emph{IEEE Transactions
  on Robotics and Automation}, vol.~19, no.~4, pp. 613--636, aug 2003.

\bibitem{Conner2005}
D.~C. Conner, A.~Greenfield, P.~N. Atkar, A.~A. Rizzi, and H.~Choset, ``{Paint
  Deposition Modeling for Trajectory Planning on Automotive Surfaces},''
  \emph{IEEE Transactions on Automation Science and Engineering}, vol.~2,
  no.~4, pp. 381--392, oct 2005.

\bibitem{Li2010a}
X.~Li, O.~A. Landsnes, H.~Chen, S.~M-V, T.~A. Fuhlbrigge, and M.~A. Rege,
  ``{Automatic Trajectory Generation for Robotic Painting Application},'' in
  \emph{Proc. 41st International Symposium on Robotics and 6th German
  Conference on Robotics}, 2010, pp. 1--6.

\bibitem{Tang2015}
Y.~Tang and W.~Chen, ``{Surface Modeling of Workpiece and Tool Trajectory
  Planning for Spray Painting Robot},'' \emph{PLoS ONE}, vol.~10, no.~5, may
  2015.

\bibitem{From2007}
\BIBentryALTinterwordspacing
P.~J. From and J.~T. Gravdahl, ``{General Solutions to functional and kinematic
  Redundancy},'' in \emph{Proc. 2007 46th IEEE Conference on Decision and
  Control}.\hskip 1em plus 0.5em minus 0.4em\relax IEEE, 2007, pp. 5779--5786.
  [Online]. Available: \url{http://ieeexplore.ieee.org/document/4434442/}
\BIBentrySTDinterwordspacing

\bibitem{From2011}
P.~J. From, J.~Gunnar, and J.~T. Gravdahl, ``{Optimal Paint Gun Orientation in
  Spray Paint Applications - Experimental Results},'' \emph{IEEE Transactions
  on Automation Science and Engineering}, vol.~8, no.~2, pp. 438--442, apr
  2011.

\bibitem{Moe2016}
\BIBentryALTinterwordspacing
S.~Moe, G.~Antonelli, A.~R. Teel, K.~Y. Pettersen, and J.~Schrimpf,
  ``{Set-Based Tasks within the Singularity-Robust Multiple Task-Priority
  Inverse Kinematics Framework: General Formulation, Stability Analysis, and
  Experimental Results},'' \emph{Frontiers in Robotics and AI}, vol.~3, no.
  April, pp. 1--18, 2016. [Online]. Available:
  \url{http://journal.frontiersin.org/article/10.3389/frobt.2016.00016}
\BIBentrySTDinterwordspacing

\bibitem{Spong2005}
M.~W. Spong and S.~Hutchinson, \emph{{Robot Modeling and Control}}.\hskip 1em
  plus 0.5em minus 0.4em\relax Wiley, 2005.

\bibitem{Golub1996}
G.~H. Golub and C.~F. {Van Loan}, \emph{{Matrix Computations}}, 3rd~ed.\hskip
  1em plus 0.5em minus 0.4em\relax Baltimore, MD: The Johns Hopkins University
  Press, 1996.

\bibitem{Chiaverini1997}
S.~Chiaverini, ``{Singularity-robust task-priority redundancy resolution for
  real-time kinematic control of robot manipulators},'' \emph{IEEE Transactions
  on Robotics and Automation}, vol.~13, no.~3, pp. 398--410, 1997.

\bibitem{Wu2014}
H.~Wu, W.~Tizzano, T.~T. Andersen, N.~A. Andersen, and O.~Ravn, ``{Hand-eye
  calibration and inverse kinematics of robot arm using neural network},''
  \emph{Advances in Intelligent Systems and Computing}, vol. 274, pp. 581--591,
  2014.

\end{thebibliography}

\end{document}